\documentclass[twocolumn, twocolappendix]{aastex63}

\shorttitle{Spiral arms and a massive dust disk with non-Keplerian kinematics}
\shortauthors{Paneque-Carreno et al.}

\begin{document}

\title{Spiral Arms and a Massive Dust Disk with non-Keplerian Kinematics:\\
        Possible Evidence for Gravitational Instability in the Disk of Elias 2-27}

\correspondingauthor{Teresa Paneque-Carreno}
\email{teresapaz.paneque@gmail.com}

\author[0000-0002-4044-8016]{T. Paneque-Carreno}
\affiliation{Departamento de Astronom\'ia, Universidad de Chile, Camino El Observatorio 1515, Las Condes, Santiago, Chile}

\author[0000-0002-1199-9564]{L. M. Pérez}
\affiliation{Departamento de Astronom\'ia, Universidad de Chile, Camino El Observatorio 1515, Las Condes, Santiago, Chile}

\author[0000-0002-7695-7605]{M. Benisty}
\affiliation{Departamento de Astronom\'ia, Universidad de Chile, Camino El Observatorio 1515, Las Condes, Santiago, Chile}
\affiliation{Unidad Mixta Internacional Franco-Chilena de Astronom\'ia, CNRS/INSU UMI 3386}
\affiliation{Univ. Grenoble Alpes, CNRS, IPAG, 38000 Grenoble, France.}

\author[0000-0002-8138-0425]{C. Hall}
\affiliation{Department of Physics and Astronomy, The University of Georgia, Athens, GA 30602, USA.}
\affiliation{Center for Simulational Physics, The University of Georgia, Athens, GA 30602, USA.}
\affiliation{School of Physics \& Astronomy, University of Leicester, University Road, Leicester, LE1 7RH, U.K.}

\author{B. Veronesi}
\affiliation{Dipartimento di Fisica, Universita degli Studi di Milano, Via Celoria, 16, Milano, I-20133, Italy}

\author[0000-0002-2357-7692]{G. Lodato}
\affiliation{Dipartimento di Fisica, Universita degli Studi di Milano, Via Celoria, 16, Milano, I-20133, Italy}

\author[0000-0002-5991-8073]{A. Sierra}
\affiliation{Departamento de Astronom\'ia, Universidad de Chile, Camino El Observatorio 1515, Las Condes, Santiago, Chile}

\author{J. M. Carpenter}
\affiliation{Joint ALMA Observatory, Avenida Alonso de C\'ordova 3107, Vitacura, Santiago, Chile}

\author[0000-0003-2253-2270]{S. M. Andrews}
\affiliation{Center for Astrophysics \textbar\ Harvard \& Smithsonian, 60 Garden St., Cambridge, MA 02138, USA}

\author[0000-0001-7258-770X]{Jaehan Bae}
\altaffiliation{NASA Hubble Fellowship Program Sagan Fellow}
\affil{Earth and Planets Laboratory, Carnegie Institution for Science, 5241 Broad Branch Road NW, Washington, DC 20015, USA}

\author[0000-0002-1493-300X]{Th. Henning}
\affiliation{Max Planck Institute for Astronomy, K\"onigstuhl 17, D-69117 Heidelberg, Germany}

\author[0000-0003-4022-4132]{W. Kwon}
\affiliation{Department of Earth Science Education, Seoul National University, 1 Gwanak-ro, Gwanak-gu, Seoul 08826, Republic of Korea}
\affiliation{Korea Astronomy and Space Science Institute, 776 Daedeokdae-ro, Yuseong-gu, Daejeon 34055, Republic of Korea}

\author[0000-0002-8115-8437]{H. Linz}
\affiliation{Max Planck Institute for Astronomy, K\"onigstuhl 17, D-69117 Heidelberg, Germany}

\author{L. Loinard}
\affiliation{Instituto de Radioastronomía y Astrofísica, Universidad Nacional Autónoma de México Morelia, 58089, México}
\affiliation{Instituto de Astronomía, Universidad Nacional Autónoma de México, Apartado Postal 70-264, Ciudad de México 04510, México}

\author[0000-0001-5907-5179]{C. Pinte}
\affiliation{School of Physics and Astronomy, Monash University, Clayton Vic 3800, Australia}
\affiliation{Univ. Grenoble Alpes, CNRS, IPAG, 38000 Grenoble, France.}

\author[0000-0001-8123-2943]{L. Ricci}
\affiliation{Department of Physics and Astronomy, California State University Northridge, 18111 Nordhoff Street, Northridge, CA 91330, USA}

\author[0000-0003-3590-5814]{M. Tazzari}
\affiliation{Institute of Astronomy, University of Cambridge, Madingley Road, CB3 0HA, Cambridge, UK}

\author{L. Testi}
\affiliation{European Southern Observatory, Karl-Schwarzschild-Strasse 2, D-85748 Garching bei M\"unchen, Germany}

\author[0000-0003-1526-7587]{D. Wilner}
\affiliation{Center for Astrophysics \textbar\ Harvard \& Smithsonian, 60 Garden St., Cambridge, MA 02138, USA}



\begin{abstract}

To determine the origin of the spiral structure observed in the dust continuum emission of Elias 2-27 we analyze multi-wavelength continuum ALMA data with a resolution of $\sim$0.2 arcsec ($\sim$23\,au) at 0.89, 1.3 and 3.3\,mm. We also study the kinematics of the disk with $^{13}$CO and C$^{18}$O ALMA observations in the $J=$3-2 transition. 
The spiral arm morphology is recovered at all wavelengths in the dust continuum observations, where we measure contrast and spectral index variations along the spiral arms and detect subtle dust-trapping signatures. We determine that the emission from the midplane is cold and interpret the optical depth results as signatures of a higher disk mass than previous constraints.
From the gas data, we search for deviations from Keplerian motion and trace the morphology of the emitting surfaces and the velocity profiles. We find an azimuthally varying emission layer height in the system, large-scale emission surrounding the disk, and strong perturbations in the channel maps, co-located with the spirals.
Additionally, we develop multigrain dust and gas SPH simulations of a gravitationally unstable disk and compare them to the observations. 
Given the large scale emission and highly perturbed gas structure, together with the comparison of continuum observations to theoretical predictions, we propose infall-triggered gravitational instabilities as origin for the observed spiral structure.

\end{abstract}



\section{Introduction}

Protoplanetary disks around young stars have shown various structures in their thermal dust continuum emission. Observing and understanding their origin is necessary to understanding the chemical, physical and dynamical processes that are ongoing in a protoplanetary disk. Of the structures present in dust emission, rings and gaps are the most common \citep[e.g., ][]{2015ApJ...808L...3A, DSHARP_Andrews, DSHARP_Huang_Annular, 2018ApJ...869...17L, 2018A&A...610A..24F}, albeit arcs and spirals have also been observed in some systems  \citep[e.g., ][]{2018ApJ...860..124D,DSHARP_Huang_Spirals}. 

Observing with instruments such as the Atacama Large Millimeter/submillimeter Array (ALMA) is crucial to our understanding of planet-formation mechanisms, as we can observe at wavelengths that trace continuum emission from the cold midplane \citep[e.g.,][]{2014prpl.conf..339T}, where we expect planets to be forming or have already formed. In the case of midplane spiral structures, their origin may be linked to the presence of a companion; stellar, fly-by or planetary \citep[]{2015MNRAS.453.1768P,2018ApJ...860L...5F, 2019MNRAS.483.4114C, 2018ApJ...859..118B, 2018ApJ...860..124D, 2020A&A...639A..62K}. Spirals may also be excited if the system is gravitationally unstable. Gravitationally instability is expected in cool and massive disks, where the disk-to-star mass ratio is larger than 0.1 \citep[][]{1997ApJ...486..372B, 2001ApJ...553..174G, 2004MNRAS.351..630L, 2016ARA&A..54..271K, 2016MNRAS.458..306H, 2016PASA...33...12R, 2020MNRAS.493.2287Z, 2019ApJ...871..228H, 2009MNRAS.393.1157C}. To date, not many spirals in dust continuum emission have a clear origin, except for those in multiple systems where the presence of spirals has been linked to stellar interactions \citep[]{DSHARP_Nico, 2020MNRAS.491.1335R}. On the other hand, there are disks were spirals have been reported at millimeter wavelengths and where no companion to which the spiral origin may be linked to has been detected yet \citep[to date these are Elias 27, IM Lup, WaOph 6, and MWC 758, ][]{laura_elias,DSHARP_Huang_Spirals, 2018ApJ...860..124D}. If no companion is detected and the disk is massive compared to the host star mass, the gravitational instability (GI) scenario arises as a possible explanation for the origin of the observed spirals. Studying disks undergoing GI is important, as population synthesis models show that GI primarily ends up forming brown dwarf mass objects \citep[]{2018MNRAS.474.5036F, 2017MNRAS.470.2517H}. It seems that giant planet formation through GI is rare \citep{2015MNRAS.454.1940R}, but it may still be the formation mechanism for important systems like HR 8799 \citep{2017A&A...603A...3V}.

Elias 2-27 is a young (0.8 Myr) M0 star \citep{2009ApJ...700.1502A} located at a distance of 116$^{+19}_{-10}$ pc \citep{2018A&A...616A...1G} in the $\rho$ Oph star-forming region \citep{1999ApJ...525..440L}. It harbors an unusually massive protoplanetary disk, the disk-to-star mass ratio of Elias 2-27 is reported to be $\sim$ 0.3 \citep[]{2009ApJ...700.1502A, laura_elias}. The initial detection of two large-scale spiral arms was obtained with medium-resolution ALMA observations by \cite{laura_elias}. Due to the brightness and accessibility of the source, it became one of the Disk Substructures at High Angular Resolution Project \citep[DSHARP,][]{DSHARP_Andrews} targets, allowing further analysis of the dust emission at high resolution. Its distinctive morphology consists of two extended quasi-symmetric spiral arms and a gap, 14\,au wide, located at 69\,au from the star \citep{DSHARP_Huang_Annular,DSHARP_Huang_Spirals}. Due to its characteristic structure, the system has been the subject of several theoretical studies, concluding that GI is a possible origin to the spiral arms \citep[]{2018MNRAS.477.1004H,2018ApJ...860L...5F, 2017ApJ...839L..24M, 2018ApJ...859..119B}. Though GI seems to explain the spiral morphology, it does not explain the dust gap, which could be carved by a companion of $\sim$0.1 M$_J$ as constrained in hydrodynamical simulations by \cite{DSHARP_Zhang}. Localized deviations from Keplerian motions at the location of this dust gap have been recently found, strengthening the hypothesis of a planetary-mass companion in the gap \citep{2020ApJ...890L...9P}. However, a lower mass inner companion, such as the one proposed to open the gap, would not be able to excite the observed spiral arms \citep{2017ApJ...839L..24M}.

Overall, Elias 2-27 appears to be a strong candidate to be a gravitationally unstable protoplanetary disk, but there are many tests to be done in order to determine if this is in fact the origin of the observed spirals. GI spirals will create pressure enhancements where we expect solids to be trapped and grain-growth favored \citep{2004MNRAS.355..543R, 2005prpl.conf.8560R, 2015MNRAS.451..974D}. This will not occur in the case of a companion, as companion-induced spirals will co-rotate with the planet at its Keplerian speed, faster than the background gas flow at their location, prohibiting dust growth and accumulation \citep{2015MNRAS.451.1147J}. Another morphological signature is the expected symmetry of the spirals produced by GI, which should have a constant pitch angle in a logarithmic spiral model \citep{2018ApJ...860L...5F}. Thus, measurements of dust growth signatures together with symmetric, constant pitch angle, logarithmic spirals, in a protoplanetary disk point towards a GI scenario. 

Additionally, valuable dynamical information may be obtained from gas observations. The presence of planets or companions leaves distinct footprints in the kinematics and these perturbations may be constrained by the amplitude of the gas deviations from the expected Keplerian motion of an unperturbed disk. The current state-of-the art methods vary from tracing pressure gradients \citep{2018ApJ...860L..12T}, observing deviations from expected isovelocity curves in the channel maps ('kinks') \citep[]{2019NatAs...3.1109P, 2018ApJ...860L..13P, 2020ApJ...890L...9P, 2015ApJ...811L...5P} and using the mean velocity maps to model the velocity structure of the disk and detect Doppler flips in the residuals \citep[]{2020ApJ...889L..24P, 2018MNRAS.480L..12P}. For GI spirals, \cite{2020arXiv200715686H} characterizes the presence of a ``GI-wiggle'' that, contrary to companion-disk interactions, will not be spatially localized but rather be a large scale perturbation, present throughout a wider velocity range, co-located with the spirals. Analyzing the disk kinematics complements the analysis of the observed dust structures and allows us to understand and connect the various ongoing processes.

Previously published gas observations of Elias 2-27 in the $^{12}$CO and $^{13}$CO in J=2-1 transition show heavy absorption, as the star is quite embedded in its cloud \citep[]{2009ApJ...700.1502A, laura_elias, DSHARP_Andrews,2020ApJ...890L...9P}. In this study we present $^{13}$CO and C$^{18}$O observations in J=3-2 transition. The higher energy transition and lower abundance of the isotopologues allows us to avoid some of the cloud contamination, while also probing closer to the midplane than previous works. 

The present paper offers new observational constraints on Elias 2-27 and is organized as follows. Section 2 provides an overview on the calibration and imaging process of the observations, section 3 analyzes the spirals in the multi-wavelength continuum data, section 4 studies the $^{13}$CO and C$^{18}$O emission, through a geometrical analysis of the moment maps and the localization of perturbations in  channel maps, section 5 shows the analysis of hydrodynamical simulations computed for a GI disk using the derived observational parameters of Elias 2-27, section 6 discusses the results and determines the possible origin of the spirals, and finally, section 7 summarizes the main findings of this work.

\section{Observations}

We present multi-wavelength (Band 3, 6 and 7) dust continuum and spectral line ($^{13}$CO $J = 3-2$ and C$^{18}$O $J = 3-2$) ALMA data of Elias 2-27. In the case of the Band 6 (1.3\,mm) observations, the imaged data corresponds to the one presented in \cite{laura_elias}, detailed information regarding the calibration of this dataset may be found in the original publication. For the Band 7 (0.89\,mm) and Band 3 (3.3\,mm) observations, the dust continuum data was first calibrated through the ALMA pipeline and afterwards, phase and amplitude self-calibration was applied. Additionally, following the calibration procedure described in \cite{DSHARP_Andrews}, we applied astrometric and flux scale alignment, to correct for spatial offsets between the center of emission of different observations and relative flux scale differences. In both bands we have short (15\,m-313.7\,m in Band 7, 15.1\,m-2.5\,km in Band 3) and long (15.1\,m-1.4\,km in Band 7, 21\,m-3.6\,km in Band 3) baseline observations, in order to account for all of the emission from different spatial scales of the disk. Spatial offsets are corrected by locating the center of emission through a Gaussian fit in the image plane of each observation and adjusting phase and pointing centers with the Common Astronomy Software Applications \citep[CASA][]{2007ASPC..376..127M} tasks \textsc{fixvis} and \textsc{fixplanets} respectively. The absolute flux calibration uncertainty of ALMA data is expected to be $\sim$10$\%$, and we note that after self-calibration the relative flux scales are consistent within $\sim$5$\%$ for different observations in the same band. To adjust these amplitude scale differences we compare the deprojected, azimuthally averaged, visibility profiles and scale them using the short-baseline data as reference for the \textsc{gaincal} task, following the methodology in \citet{DSHARP_Andrews}. For Band 7 data we apply the spatial offset corrections and amplitude scaling before any self calibration as relative flux scales vary $5-10\%$ between observation sets. In the Band 3 data set we have a higher flux scale difference from the initial datasets ($\sim$20$\%$), probably due to atmospheric conditions as we can visually see that the image quality is not the best. Therefore, in Band 3 we conduct self-cal of the short-baseline observation and afterwards, with a flux scale difference of $\sim$3$\%$, we perform the spatial and amplitude scale corrections.

Self-calibration was first conducted on the dust continuum emission short-baseline data and the result combined with the long-baseline data to be self-calibrated all together. The selected time intervals for both phase and amplitude self-calibration start with the longest available total integration period for the initial round, and afterwards, each following time interval was half of the previous interval. The longest and shortest intervals used were, 1500s and 46s for the phase calibration of the joint 0.89\,mm data, and 2860s and 178s for the phase calibration of the joint 3.3\,mm data. For the calibration of the short baseline data, the time intervals of phase calibration were between 1080s and 270s for the 0.89\,mm data, and 304s and 9.5s for the 3.3\,mm data. In the case of the amplitude self-cal time intervals, in Band 7 only one round was required in both short baseline and joint data, so the solutions were obtained from the longest time interval. In Band 3 the short baseline data had one round of amplitude calibration and the joint data set had two rounds. Self-calibration rounds were applied until signal-to-noise ratio (SNR) improvement was less than 5$\%$ in the case of Band 7 data and until there was no SNR improvement in the case of Band 3.  For the short baseline data at 0.89\,mm, we conducted 3 rounds of phase calibration and 1 round of amplitude calibration. The combined dataset at 0.89\,mm had 7 rounds of phase calibration and 1 round of amplitude calibration, overall the peak SNR in the joint dataset improved 300$\%$. The short baseline data at 3.3\,mm had 6 rounds of phase calibration and 1 round of amplitude calibration, the joint data set had 5 rounds of phase calibration and 2 round of amplitude calibration. In the joint dataset, the SNR improvement was 6$\%$.

For the final imaging we used multi-scale \textsc{tclean} for all the images, using a 1$\sigma$ stopping threshold (where $\sigma$ is the image RMS) for the Bands 3 and 6 data, and a 2$\sigma$ stopping threshold in Band 7. Robust weighting values were 1.0 in the case of Band 6, and 0.5 for Bands 3 and 7. We also set the \textsc{gain} value to 0.05 and \textsc{cyclefactor} parameter in \textsc{tclean} to 2.0, to have a more detailed cleaning, by substracting a smaller fraction of the source flux from the residual image and triggering a major cycle sooner. All images have comparable beam sizes: 0.22$\arcsec \times$0.17$\arcsec$ beam (26\,au$\times$20\,au) at 0.89\,mm, 0.26$\arcsec \times$0.22$\arcsec$ beam (30\,au$\times$26\,au) at 1.3\,mm, and 0.26$\arcsec \times$0.20$\arcsec$ (30\,au$\times$23\,au) beam at 3.3\,mm. At each wavelength the image RMS is approximately 93\,$\mu$Jy/beam, 86\,$\mu$Jy/beam and 10\,$\mu$Jy/beam, respectively for 0.89\,mm, 1.3\,mm and 3.3\,mm. 

The observations of C$^{18}$O and $^{13}$CO were obtained simultaneous to the dust continuum emission in Band 7 (0.89\,mm) and in the $J = 3 - 2$ transition. C$^{18}$O is observed at 329.330 GHz, with a spectral resolution of 0.035 MHz and $^{13}$CO at 330.588 GHz, with a spectral resolution of 0.121 MHz. After applying the same self-calibration solutions as to the dust continuum of Band 7, the emission was first imaged using natural weighting (robust parameter of 2.0) and not applying any uvtapering or uv-range filtering, in order to be sensitive to large-scale emission with a beamsize of 0.29$\arcsec \times$0.22$\arcsec$ ($\sim$ 34$\times$25\,au) for $^{13}$CO and 0.30$\arcsec \times$0.23$\arcsec$ ($\sim$ 35$\times$27\,au) for C$^{18}$O (see figures in Appendix A). In order to avoid the cloud contamination present in the disk \citep[]{laura_elias, DSHARP_Huang_Spirals} we exclude large-scale emission by considering only baselines longer than 36m for $^{13}$CO (scales shorter than 6.4$\arcsec$, $\sim$740\,au) and 45m for C$^{18}$O (scales shorter than 5.11$\arcsec$, $\sim$590\,au). We also applied uvtapering of 0.2$\arcsec \times$0.115$\arcsec$, PA= 0$^{\circ}$ for $^{13}$CO and 0.2$\arcsec \times$0.0$\arcsec$, PA= 0$^{\circ}$ for C$^{18}$O, in order to obtain a roughly round beam. The final images that trace the disk emission were  obtained with a robust parameter of 0.5, resulting in a beam size of 0.26$\arcsec \times$0.25$\arcsec$ ($\sim$ 29\,au) for $^{13}$CO and 0.31$\arcsec \times$0.29$\arcsec$ ($\sim$ 35\,au) for C$^{18}$O. We imaged all channel maps in both $^{13}$CO and C$^{18}$O using a 0.111\,km s$^{-1}$ spectral resolution, though the C$^{18}$O data was observed with a finer spectral resolution (0.032\,km s$^{-1}$), and found the best compromise of SNR by using a broader spectral resolution.

We also recover gas emission from CN v$ = 0$, N $= 3 - 2$ in $J = 7/2 - 5/2$ and $J = 5/2 - 3/2$, these additional molecules will be studied separately in future work. We do not recover any emission from requested observations of CN v$ = 0$, N $= 3 - 2$ in $J = 5/2 - 5/2$ or SO 3$\Sigma$ v$=0$ $J = 3 - 2$, at the achieved sensitivity level of 3mJy/beam. 

\begin{figure*}
   \centering
   \includegraphics[width=\hsize]{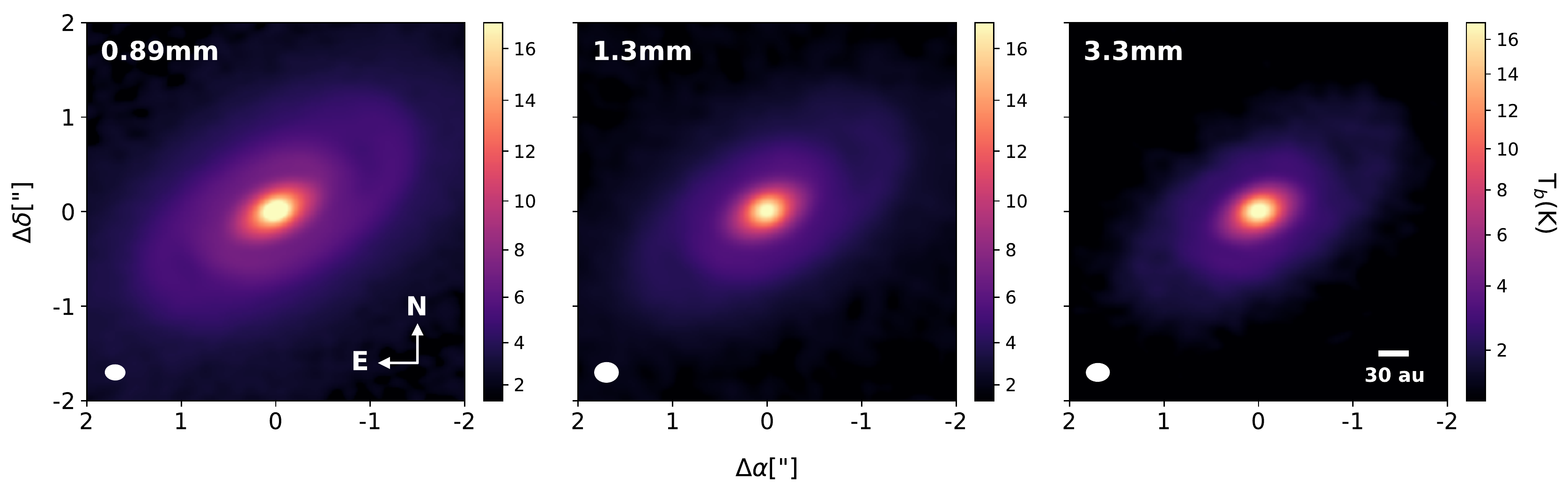}
      \caption{Dust continuum observations of Elias 2-27 at 0.89\,mm, 1.3\,mm and 3.3\,mm. For each panel: the intensity colorscale is shown on the right, the scalebar in lower right corner corresponds to 30\,au at the distance of the star, and the ellipse in the bottom left corner indicates the spatial resolution.
              }
         \label{dust_data}
\end{figure*}

\section{Dust Spiral Structure}

We recover the spiral structure at all wavelengths, as shown in Figure \ref{dust_data}. Given the $\sim$25\,au beam size, we are not able to fully resolve the 69\,au gap but can distinguish it as a small decrease in brightness temperature at all wavelengths in Figure \ref{dust_data}.
For all calculations throughout this study, we will assume the gap location, disk inclination, and disk position angle as derived in \cite{DSHARP_Huang_Annular}, which are 69.1 $\pm 0.4$\,au, 56.2$^\circ \pm 0.8 ^\circ$, and 118.8$^\circ \pm 0.7 ^\circ$, respectively. 

\begin{figure*}
   \centering
   \includegraphics[width=\hsize]{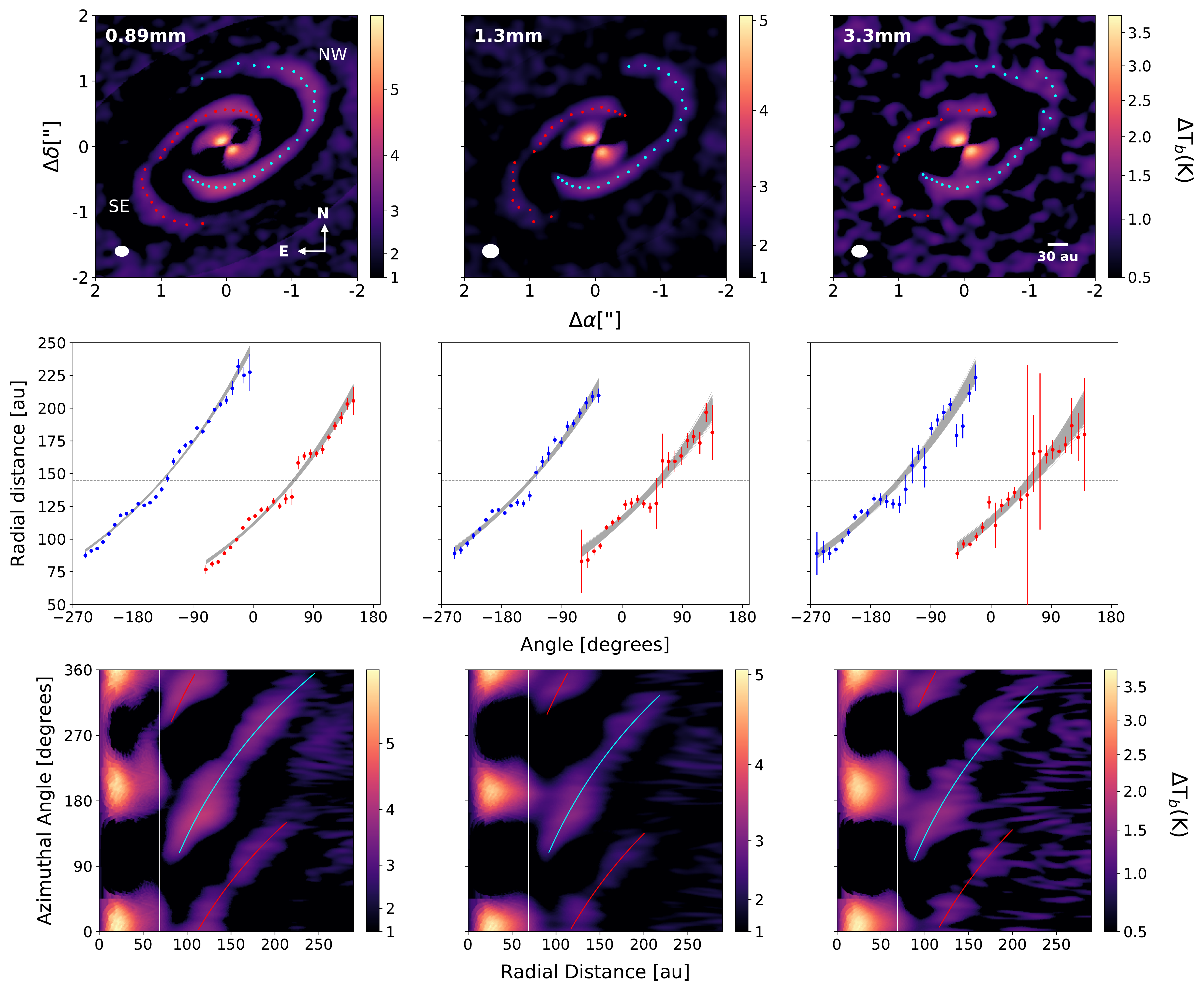}
      \caption{ 
     The spiral morphology of Elias 2-27 at multiple wavelengths. Panels from left to right correspond to data from the 0.89\,mm, 1.3\,mm and 3.3\,mm observations. North-West spiral is traced in blue, South-East spiral is traced in red. Top panels: Dust continuum maps from which the azimuthally averaged radial profile has been subtracted to highlight spiral location, red and blue points trace the maxima of emission along the arms. Middle panels: Deprojected radial location of the emission maxima, as a function of the angle measured from North to the East (left). Error bars correspond to the astrometric error of each data point and grey lines show the posterior distribution of a logarithmic spiral fit with constant pitch angle. Bottom panels: Deprojection of the subtracted dust continuum observations from the top panels. The vertical line marks the dust gap location from \citet{DSHARP_Huang_Spirals}, and colored lines show the best-fit logarithmic spiral model.
              }
         \label{spiral_trace}
\end{figure*}

\begin{table*}
\def\arraystretch{1.5}
\caption{Best-fit Parameters of the Logarithmic Spiral Model}
\label{table_dust_pitch}      
\centering                          
\begin{tabular}{c c c c c c}       
\hline\hline                
Wavelength & Spiral Arm & Angle Range & $R_0$ [au] & $b$ & Pitch Angle\\    
\hline                       
            
     0.89\,mm&      NW & -251$^{\circ}$ to -5$^{\circ}$ & 249.8$^{+1.2}_{-1.1}$ & $0.230\pm0.002$ & 12.9$^{\circ}$ $\pm 0.1^{\circ}$\\
     
        &       SE & -70$^{\circ}$ to 150$^{\circ}$ & 111.5$\pm0.4$ & $0.247\pm0.003$ & 13.9$^{\circ} \pm 0.2^{\circ}$\\
        
     1.3\,mm&      NW & -250$^{\circ}$ to -35$^{\circ}$ & 250.2$^{+2.3}_{-2.5}$ & $0.229^{+0.003}_{-0.004}$ & 12.9$^{\circ} \pm 0.2^{\circ}$\\
     
        &       SE & -61$^{\circ}$ to 135$^{\circ}$ & 115.5$^{+0.8}_{-0.9}$ & 0.234$\pm0.007$ & 13.2$^{\circ} \pm 0.4^{\circ}$\\
        
     3.3\,mm&      NW & -260$^{\circ}$ to -23$^{\circ}$ & 249.9$^{+3.9}_{-3.8}$ & $0.229\pm0.005$ & 12.9$^{\circ} \pm 0.3^{\circ}$\\
     
        &       SE & -50$^{\circ}$ to 140$^{\circ}$ & 113.6$^{+1.0}_{-1.1}$ & $0.231\pm 0.010$ & 13.0$^{\circ} \pm 0.6^{\circ}$\\

\hline                                   
\end{tabular}
\end{table*}

\subsection{Tracing the Spiral Morphology}
To trace the spiral structure from our dust continuum images, we radially subtract an azimuthally-averaged radial profile of the emission and trace the spiral features from these ``subtracted images'' \citep[as done in ][]{DSHARP_Huang_Spirals}. From the subtracted images, shown in the top panels of Figure \ref{spiral_trace}, we find the radial location of maximum emission along each spiral, at azimuthal steps sampled every 9$^{\circ}$, between a determined azimuthal angle range (values in Table \ref{table_dust_pitch}). The radial extent where we trace the maxima of emission is determined visually. To aid our visual criteria we consider radial locations no further than where we have a signal of 5 times the RMS in the non-subtracted image. 

Previous analyses of Elias 2-27 \citep[][]{laura_elias, DSHARP_Huang_Spirals} have shown that a logarithmic spiral model, with a constant pitch angle, can adequately trace the spiral morphology. Therefore, we use MCMC modelling to find the best-fit parameters for a logarithmic spiral model, considering the location data of each spiral and each observed wavelength. The spiral form is given by:

\begin{equation}
    r(\theta) = R_0 e^{b\theta}
\end{equation}

Here $\theta$ is the polar angle in radians measured from the North and to the East (left), for reference see coordinates in Figure \ref{dust_data}. $R_0, b$ are free parameters, with $R_0$ the radius at $\theta = 0$, measured in au, and $b$ relates to the pitch angle of the spiral arms ($\phi$), as $\phi = arctan(b)$. The uncertainty on the location of each measured maxima is assumed to be the astrometric error\footnote{see Sect. 10.6.6 in \url{https://almascience.nrao.edu/documents-and-tools/cycle5/alma-technical-handbook/view} for further details.}: $\Delta p = 60000 \cdot (\nu \cdot B \cdot SNR)^{-1}$, were $\Delta p$ is the approximate position uncertainty of a feature in milliarcseconds, $SNR$ is the peak/RMS intensity ratio of the data point on the image, $\nu$ is the observing frequency in GHz and $B$ is the maximum baseline length in kilometers.

The top panel of Figure \ref{spiral_trace} shows the maxima along each spiral from the subtracted images. In the middle panels are the deprojected radial locations of each maxima, measured from the center, as a function of azimuthal angle. Grey lines show logarithmic spiral models, derived from 300 draws of the posterior values after convergence of the MCMC simulations for the best-fit parameters of each spiral arm. The horizontal line at 148\,au marks the location where we observe a break in the spiral arm, at all wavelengths, clearest in the South-East spiral, also subtly present in the North-West spiral. \cite{DSHARP_Huang_Spirals} had previously noted a possible decrease in the pitch angle value outside R $\sim$150\,au. The bottom panel (Figure \ref{spiral_trace}) shows the polar deprojection of the subtracted data, with a vertical line marking the dust gap location. The red/blue lines show the best-fit model for the logarithmic spirals. The median value for the parameters of the logarithmic spiral model, for each spiral and wavelength, are shown in Table \ref{table_dust_pitch}, along with the 16th and 84th percentile uncertainties derived from the posteriors. 

We note that the pitch angle values retrieved here, $\sim$ 12.9$^{\circ}$ and $\sim$ 13.3$^{\circ}$, for NW and SE spiral respectively (similar between wavelengths, see Table \ref{table_dust_pitch}), are different than the recovered pitch angles from \cite{DSHARP_Huang_Spirals} (15.7$^{\circ}$ and 16.4$^{\circ}$, for NW and SE spiral). When applying our method to the high-resolution dataset presented in \cite{DSHARP_Huang_Spirals}, we retrieve the same results as in their work. The pitch angle difference between this work and theirs is probably due to beam smearing effects, combined with the challenge of image subtraction in lower-resolution data \citep[as discussed in][]{DSHARP_Huang_Spirals}, as the angular resolution difference between datasets is a factor of 4-5.

\subsection{Contrast variations along the spirals}

\begin{figure*}[t]
   \centering
   \includegraphics[width=\hsize]{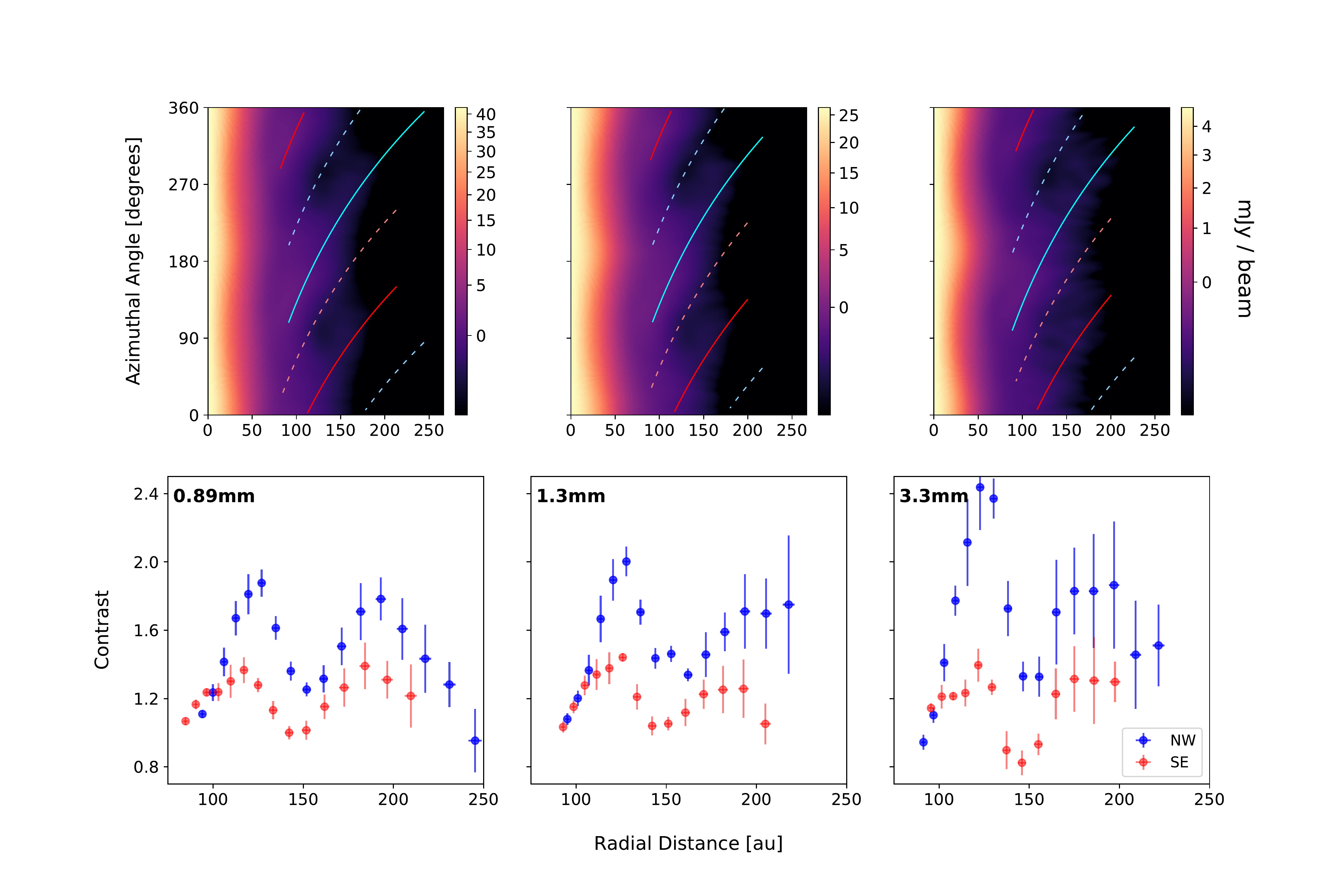}
      \caption{Contrast of the spiral arms in Elias 2-27, panels from left to right correspond to data from the 0.89\,mm, 1.3\,mm and 3.3\,mm observations. Top panels: Polar deprojection of the dust continuum emission maps, where the azimuthal angle is measured from North and to the East. Blue and red colors correspond to the North-West and South-East spirals, respectively. Continuous lines trace the spirals following the best-fit parameters found in section 3.1, dashed lines trace the inter-arm region following the same best-fit parametric model, but minus 90$^{\circ}$ from the spiral. Bottom panels: Calculated contrast values along each spiral arm, blue and red points correspond to the North-West and South-East spirals, respectively.
              }
         \label{spiral_contrast}
\end{figure*}

We test the contrast at comparable locations over different wavelengths, for this may provide evidence of dust trapping. If indeed there is dust growth within the spiral arm or large particles are trapped at this location, then at longer wavelengths we could expect to observe higher contrast. The latter effect is due to larger particles being more densely packed in the spiral arm \citep{2004MNRAS.355..543R}, this has been shown to occur in the case of dust-trapping vortex studies \citep{2018A&A...619A.161C}.  We compute how the contrast varies radially along each spiral arm, with respect to a fixed ``inter-arm'' region, which is expected to be a location where we can be sure there is no emission from the spiral arms. Considering the symmetry between the spirals, the inter-arm region of each spiral is assumed to follow the same shape of the corresponding spiral arm, but rotated by 90 degrees clockwise from the spiral location. 
The top panels of Figure \ref{spiral_contrast} show a polar projection of the dust continuum, overlayed are the best-fit model of the spirals for each wavelength in continuous line and this same model, shifted 90 degrees clockwise, tracing the inter-arm region, in dashes. From this figure it is clear that the interarm region effectively traces zones of lower emission, compared to the spiral arm location.

The contrast is calculated, from the original images, as the ratio between the emission at the spiral and the inter-arm regions, at each radius and for both spiral arms. We use the previously derived parameters and calculate the contrast along the best-fit spiral model for each wavelength, binning the data azimuthally every 15$^{\circ}$. The standard deviation within the bin is used as the error of each averaged flux measurement, both for the interarm and spiral region. The contrast curves for each wavelength are shown in the bottom panels of Figure \ref{spiral_contrast}. The NW spiral remains the strongest spiral at all wavelengths with an average contrast value 25-33$\%$ higher than the SE spiral. Both spirals maintain a similar contrast throughout their radial extent, deviating in the 0.89\,mm and 1.3\,mm observations only $\sim$11$\%$ from the average contrast value in the case of the SE spiral and $\sim$17$\%$ in the case of the NW spiral. In the 3.3\,mm data the contrast variation increases and deviations from the average value increase to $\sim$16$\%$ in the SE spiral and $\sim$24$\%$ in the NW. For the 0.89\,mm, 1.3\,mm and 3.3\,mm emission, the radial distance of the maximum contrast value is, respectively for (NW, SE) spirals, ($\sim$125\,au, $\sim$117\,au), ($\sim$129\,au, $\sim$127\,au) and ($\sim$129\,au, $\sim$126\,au). For the minimum contrast location, in the same order, the radial distances are ($\sim$155\,au, $\sim$149\,au), ($\sim$161\,au, $\sim$147\,au) and ($\sim$150\,au, $\sim$142\,au).  We observe in our contrast curves that the radial location of the maximum contrast slightly shifts outwards with wavelength, no global shifts are detected in the minimum contrast distances, however the minimum contrast of the SE spiral appears to shift inwards at longer wavelength. Overall our values are in agreement with previous measurements of \cite{DSHARP_Huang_Spirals} where the peak contrast is located at $\sim$123\,au and low contrast at $\sim$147\,au (measured from higher angular resolution 1.3\,mm data). While the form of the contrast curve in the 1.3\,mm emission agrees also with the previous studies \citep[]{laura_elias, DSHARP_Huang_Spirals}, at larger radial distances we observe a decrease in the contrast of both spirals in the 0.89\,mm emission and also in the SE spiral in 3.3\,mm, contrary to the apparent growing contrast at larger radial distance observed in the 1.3\,mm emission.

If we follow the peak contrast of both spirals, located at $\sim$125\,au, we see a slight variation in the contrast value, which grows larger with longer wavelength. The peak contrast value of the NW spiral increases by 11$\%$ from the 0.89\,mm to the 1.3\,mm emission and by 8$\%$ between the 1.3\,mm and 3.3\,mm emission. The SE spiral increases its peak contrast value 5$\%$ initially and in the 3.3\,mm data decreases the peak contrast by 2$\%$ with respect to the 1.3\,mm emission. Minimum contrast values also shift, but do not show any correlation with varying wavelength. We note that previous studies on this source measured the contrast differently, using the ratio between the spiral flux and the minimum flux at the same radial location \citep[]{DSHARP_Huang_Spirals, laura_elias}. For completeness, we test our results using the minimum flux method used in the previous studies, and obtain similar contrast curves in which we also measure increasing peak contrast values towards longer wavelengths. However, we do not consider these results, for the minimum flux at a determined radial distance is found at different azimuthal angles between different wavelengths. Therefore, the contrast calculated using the minimum radial flux does not sample comparable locations between different wavelengths, something necessary in order to detect possible signatures of dust growth at a given location, which we achieve by using the interarm region. 

\subsection{Spectral Index Analysis}

\begin{figure}[h!]
   \centering
   \includegraphics[width=\hsize]{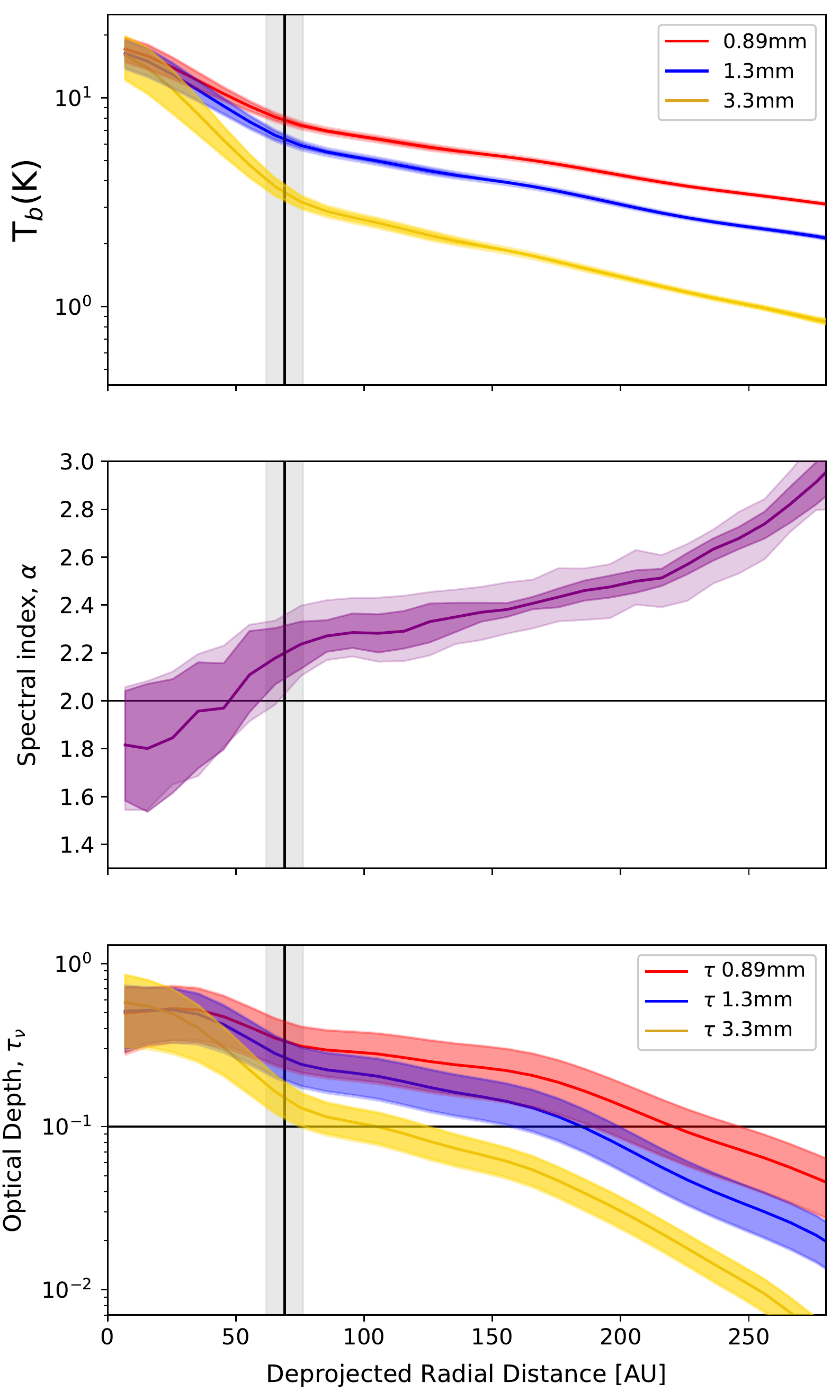}
      \caption{Radial profiles of brightness temperature, spectral index, and optical depth for Elias 2-27 using multiple wavelengths. Top panel: Azimuthally averaged brightness temperature profiles for each wavelength, shaded area shows the 1$\sigma$ scatter at each radial bin divided by the beams spanning the angles over which the intensities are measured. Middle panel: Spectral index radial profile considering all three wavelengths, darker shaded area shows 1$\sigma$ errors from assuming the intensity error of the azimuthally averaged intensity profile. Lightly shaded area shows the total error including a 10\% flux uncertainty on the intensity measurement. Lower panel: Optical depth radial profile for each wavelength, shaded area shows 1$\sigma$ uncertainty considering the 1$\sigma$ errors on the stellar luminosity and from the azimuthally averaged intensity profiles. Vertical dashed line and shaded region marks the dust gap location (69\,au) and width (14.3\,au) as determined in \cite{DSHARP_Huang_Annular}.
              }
         \label{radial_prof_tau_alpha}
\end{figure}

\begin{figure*}
   \centering
   \includegraphics[width=\hsize]{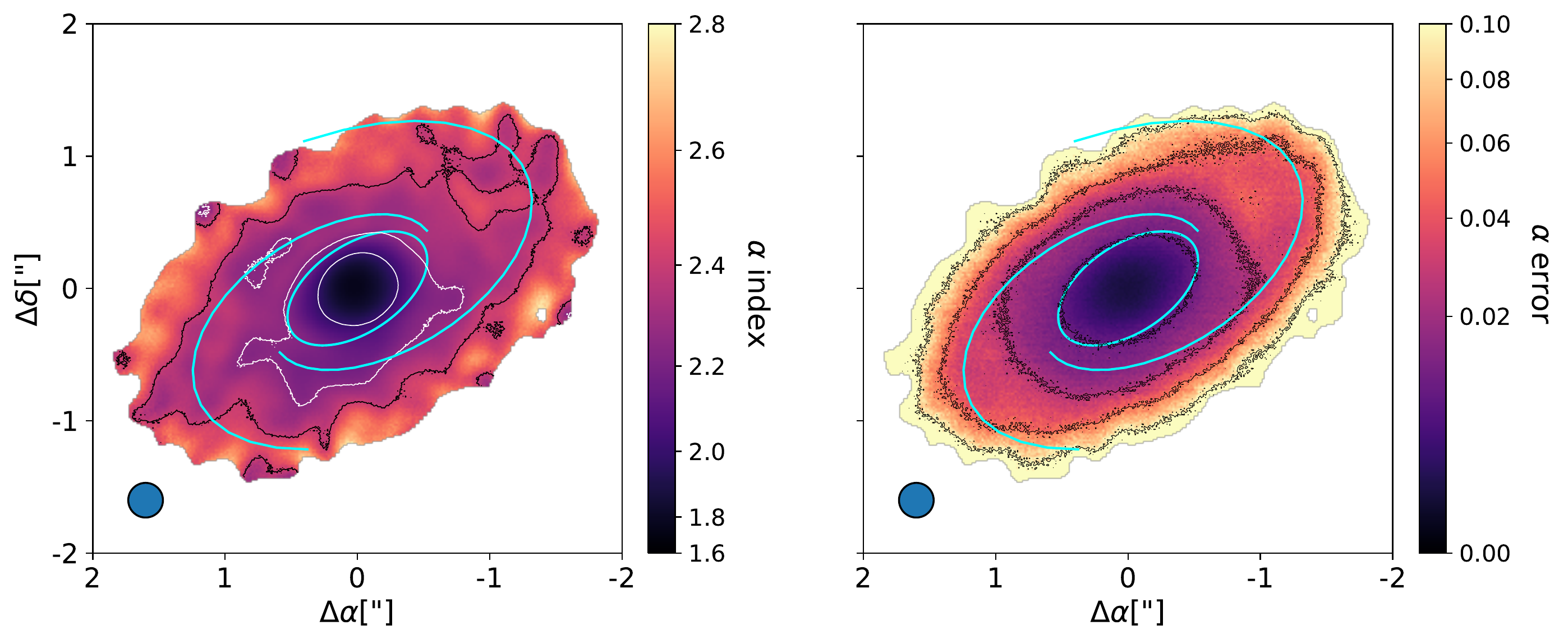}
      \caption{Left: Spectral index map calculated from the emission of all available wavelengths. Only emission over 5$\sigma$ in the image plane is considered. Blue lines show the location of the dust gap at 69\,au and the derived best-fit parametric model for the spiral arms of 1.3\,mm emission. White contours correspond to $\alpha$=2.0 and 2.25, black contours indicate $\alpha$=2.4. Right: Spectral index error map, obtained from the image rms and intensity value of each pixel. Blue lines trace the dust features and black contours correspond to $\alpha$ uncertainties at 0.01, 0.025, 0.05 and 0.1 levels. 
              }
         \label{spectral_index_map}
\end{figure*}

The presence of dust growth throughout the disk can be inferred using the multi-wavelength continuum observations to compute the disk spectral index. The spectral index ($\alpha$) of the spectral energy distribution follows $I_{\nu} \propto \nu^{\alpha}$, where $I_{\nu}$ corresponds to the measured intensity, from the image plane, at frequency $\nu$. In the optically thin regime, if $\alpha$ has values between 2-2.5 it can be an indicator towards the presence of large, up to cm-sized, grains and hence dust growth at the location of such values \citep[e.g., ][]{2009ApJ...696..841K, 2014prpl.conf..339T, 2017ApJ...839...99P, 2018A&A...619A.161C, 2019ApJ...881..159M}. To adequately compare the emissions from all three wavelengths, the calculations in this section are done from dust images at equal resolution and centered in the peak brightness. We use the task imsmooth in CASA to especify a round, 0.26$\arcsec \times$0.26$\arcsec$ beam ($\sim$31\,au). Figure \ref{radial_prof_tau_alpha} shows in the top panel the azimuthally averaged brightness temperature of the smoothed, equal resolution images.

We calculate $\alpha$ as the best-fit slope of a linear model applied to the logarithmic space ($\alpha \propto d$\,ln$I_{\nu}/d$\,ln$\nu$), using all wavelengths available in this study. From an MCMC fit we obtain a posterior distribution of $\alpha$, with best-fit and uncertainties computed from the 50th, 16th, and 84th percentile values.
From the azimuthally averaged intensity profiles, we compute an $\alpha$ radial profile shown in the middle panel of Figure \ref{radial_prof_tau_alpha}. We retrieve a disk average $\alpha$ value of 2.6$\pm$0.06, with $\alpha<2.0$ in the inner $\sim$50\,au. These low values in the inner disk will be discussed in section 5.1 and are probably related to optically thick emission where self-scattering at long wavelengths is a relevant process.

We apply the same techniques described in \cite{DSHARP_Huang_Annular} to calculate the optical depth at our various wavelengths. The main assumption relies on approximating the midplane temperature profile ($T_{mid}$), assuming a passively heated, flared disk in radiative equilibrium \citep[e.g.,][]{1997ApJ...490..368C} and considering that the millimeter emission follows this temperature profile, if it traces the midplane. From the temperature profile, we compute the expected blackbody emission ($B_\nu$), using Planck's law, and relate it to the measured intensity ($I_\nu$) through $\tau_\nu$, following $I_\nu (r) = B_\nu(T_{mid}(r))(1 - \exp(-\tau_\nu (r)))$. $T_{mid}$ will depend on the assumed flaring of the disk and the stellar luminosity \citep[we use the DSHARP values for flaring and stellar luminosity of $\varphi$ = 0.02 and $log L_*/L_\odot$ = -0.04 $\pm$ 0.23,][]{DSHARP_Andrews}.

\begin{figure*}
   \centering
   \includegraphics[width=\hsize]{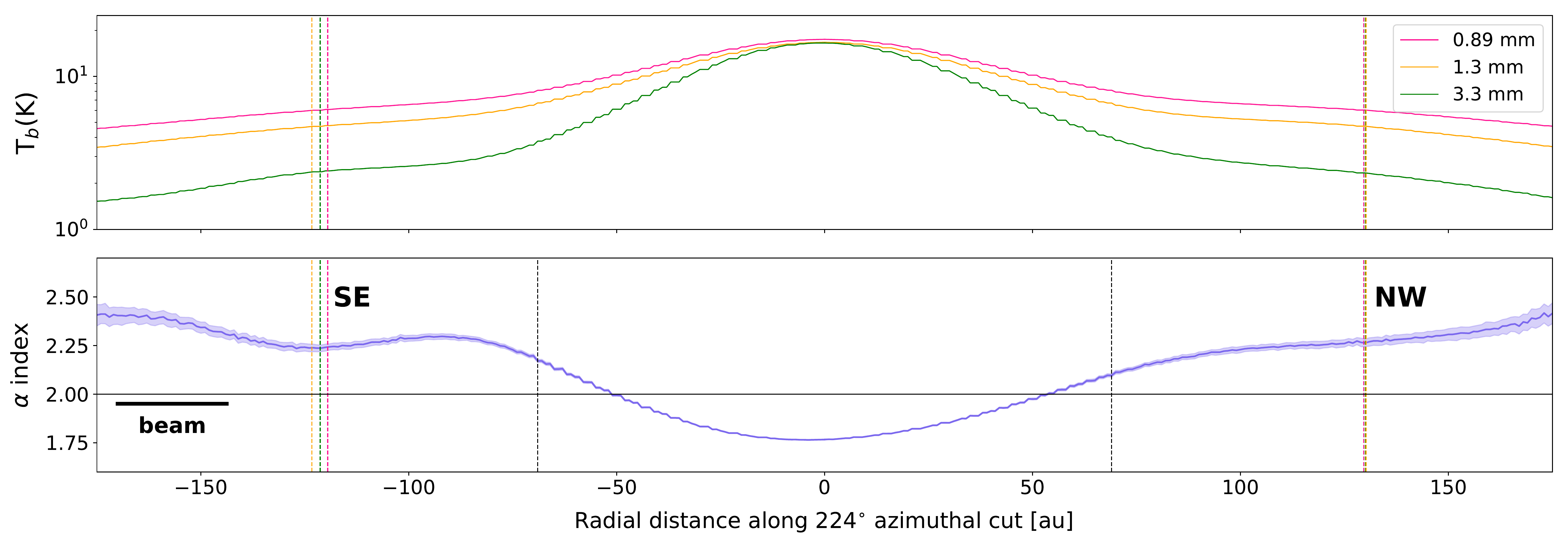}
      \caption{Top: Brightness temperature profile for 0.89\,mm, 1.3\,mm and 3\,mm emission along the azimuthal angle of maximum spiral contrast (224$^\circ$), measured radially. Positive radial values are measured from the center of the disk and to the West, negative values indicate distance to the East. Vertical coloured dashed lines show the location of the spiral arm according to the best-fit parametric model at each wavelength, black vertical line marks the location of the dust gap. Bottom: spectral index along the azimuthal cut, calculated from the emission of the three wavelengths. 
              }
         \label{panel_spiral_width}
\end{figure*}

The computed optical depth profiles are shown in the bottom panel of Figure \ref{radial_prof_tau_alpha}. We note that at 0.89\,mm the modeled $T_{mid}$ is $\sim$2 times larger than the measured $T_b$. In Rayleigh-Jeans regime, as expected for mm emission, $T_b = T_{mid}(1 - $exp$(-\tau))$. Then, if $T_{mid}$ is overestimated it will produce lower optical depth values and the emission will appear more optically thin than it really is. The disk could be colder, or with a lower flaring value, and the emission at all wavelengths would be more optically thick. These issues will be discussed in section 5.1.

The spectral index map for Elias 2-27 and the uncertainty on the spectral index are presented in Figure \ref{spectral_index_map}. We only consider emission above 5$\sigma$ at all wavelengths for this calculation, and we adopt the rms of each image as the uncertainty on the intensity for each pixel. The best-fit spiral model from the 0.89\,mm dust emission is overlaid for reference, along with the dust gap location (69\,au). Contour lines for $\alpha$ values of [2.0, 2.25, 2.4] and for $\alpha$ uncertainties at the level of [0.01, 0.025, 0.05, 0.1] are presented. We observe that contours tracing $\alpha=$ 2.4 seem to be coherent with the spiral morphology as traced by the best-fit spiral. This is specially apparent in the NW spiral location. 
The variations between the spiral and inter-arm region at the NW spiral location are $\sim0.1$, with uncertainties of $\sim0.04$. Overall, we observe lower $\alpha$ co-located with the spiral features. At the gap location no particular behaviour is observed, except for a decrease in $\alpha$ values inwards from the gap location. As noticed in the radial profile, in the inner disk ($<$50\,au) $\alpha$ reaches values below 2.0.

\begin{figure*}
   \centering
   \includegraphics[scale=0.8]{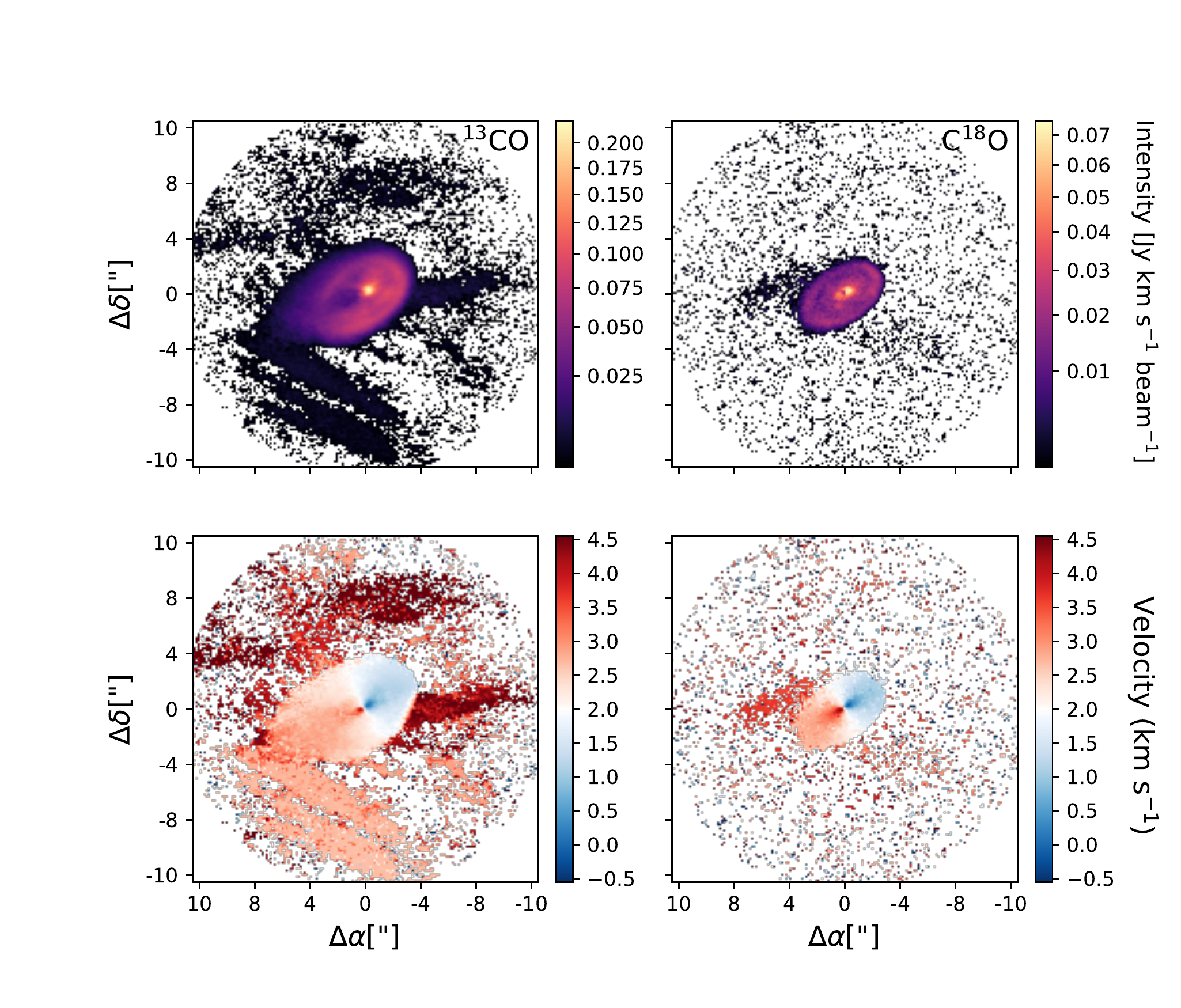}
      \caption{Left column shows $^{13}$CO emission maps, Right row C$^{18}$O emission maps. Top Row: Integrated emission maps considering all emission over 3$\sigma$. Lower Row: Mean velocity maps.
              }
         \label{mom_largescale}
\end{figure*} 

Another indicator of dust growth is that smaller grains, traced by shorter wavelengths, will be less concentrated than larger grains observed at longer wavelengths. This will translate into a width difference along the dust-trapping structure, such that smaller grains will be more widely spread than larger grains within a dust-trap. Such behaviour has been constrained in different sources with vortex-like structures, likely tracing dust traps \citep[]{2015ApJ...812..126C, 2015ApJ...810L...7V, 2018A&A...619A.161C, 2019MNRAS.483.3278C}. The measurement of the width variation cannot be done using a subtracted image or an azimuthally averaged profile, both options would introduce artifacts or remove relevant emission, therefore, we must use the original image. We decide to trace variations along a single azimuthal cut, choosing the angle were the highest spiral contrast is observed, which is located at $\sim$125\,au corresponding to $\sim$224$^\circ$.

Figure \ref{panel_spiral_width} shows the brightness temperature profiles along the azimuthal angle of maximum spiral contrast (224$^\circ$) for all three wavelengths, the spirals can be localized as a bump in the intensity curve between 100\,au-150\,au in the top panel. From this it can be seen that the shape of the intensity curve is similar at all three wavelengths, with no noticeable width differences. The bottom panel shows the spectral index distribution along the azimuthal cut, calculated directly from the top panel intensity profile, as previously done for the spectral index map. In this $\alpha$ profile we see that the East side (negative radial distance) shows a small decrease in the spectral index value at the spiral location, while the West spiral does not show significant $\alpha$ variations. The spatial extent of the variation in the East is resolved by our beam size (0.26$\arcsec$, $\sim$31\,au). 

   
\section{$^{13}$CO and C$^{18}$O $J=3-2$ emission analysis}

\begin{figure*}
   \centering
   \includegraphics[width=\hsize]{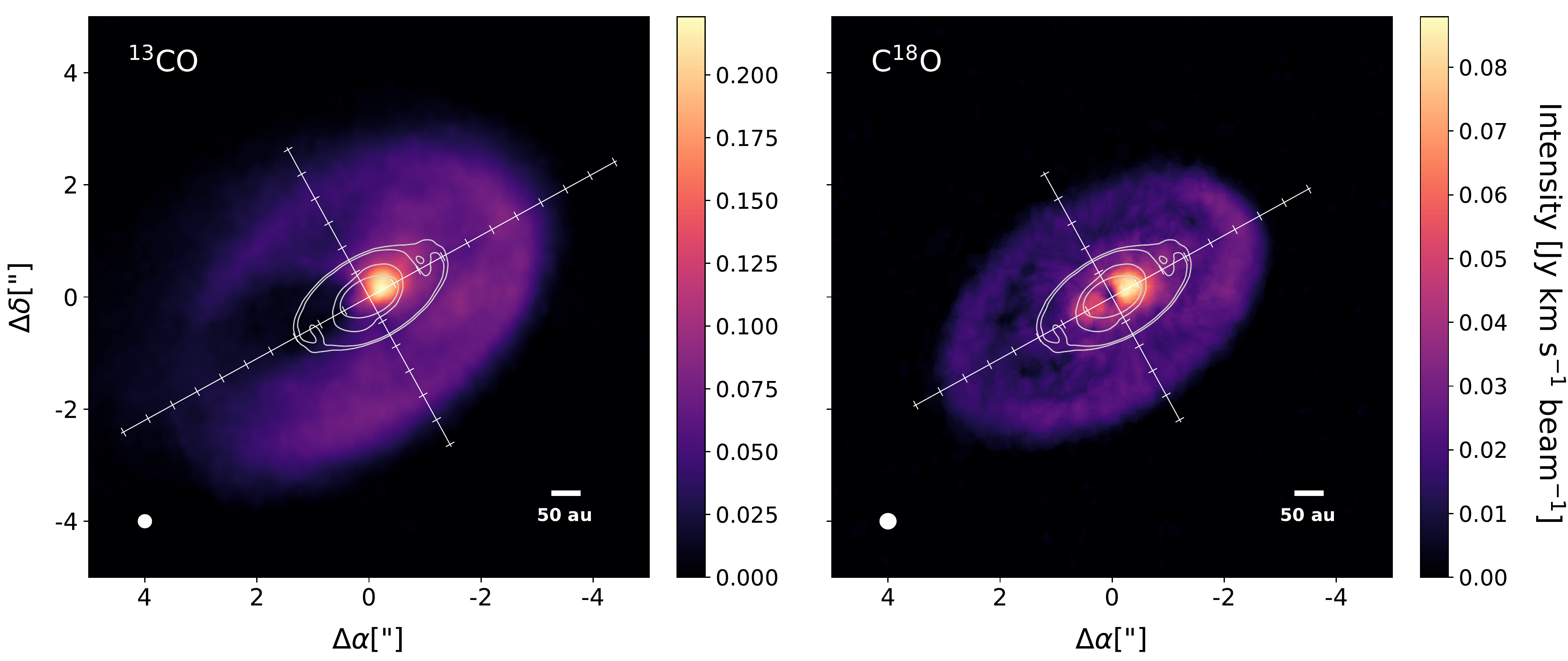}
      \caption{ Integrated emission (moment 0) maps for $^{13}$CO (left) and C$^{18}$O (right) gas emission. Contours of 0.89\,mm continuum emission are overlaid on top.  White grid marks the minor and major axis of the disk, as determined by the continuum emission position angle, with ticks on these axis indicate 0.5$\arcsec$ ($\sim$58\,au) intervals.
              }
         \label{mom0_size}
\end{figure*}

We analyze the emission from the two observed CO isotopologues in three different ways, by: 1) studying the presence of structures in the integrated emission and in channel maps, including large-scale emission in the entire field-of-view (FOV), 2) searching for velocity perturbations in channel and velocity maps, 3) constraining the vertical height of the $^{13}$CO and C$^{18}$O emitting layer, and analyzing the kinematics of these isotopologues.

\subsection{Channel and moment maps}

\begin{figure}
   \centering
   \includegraphics[width=\hsize]{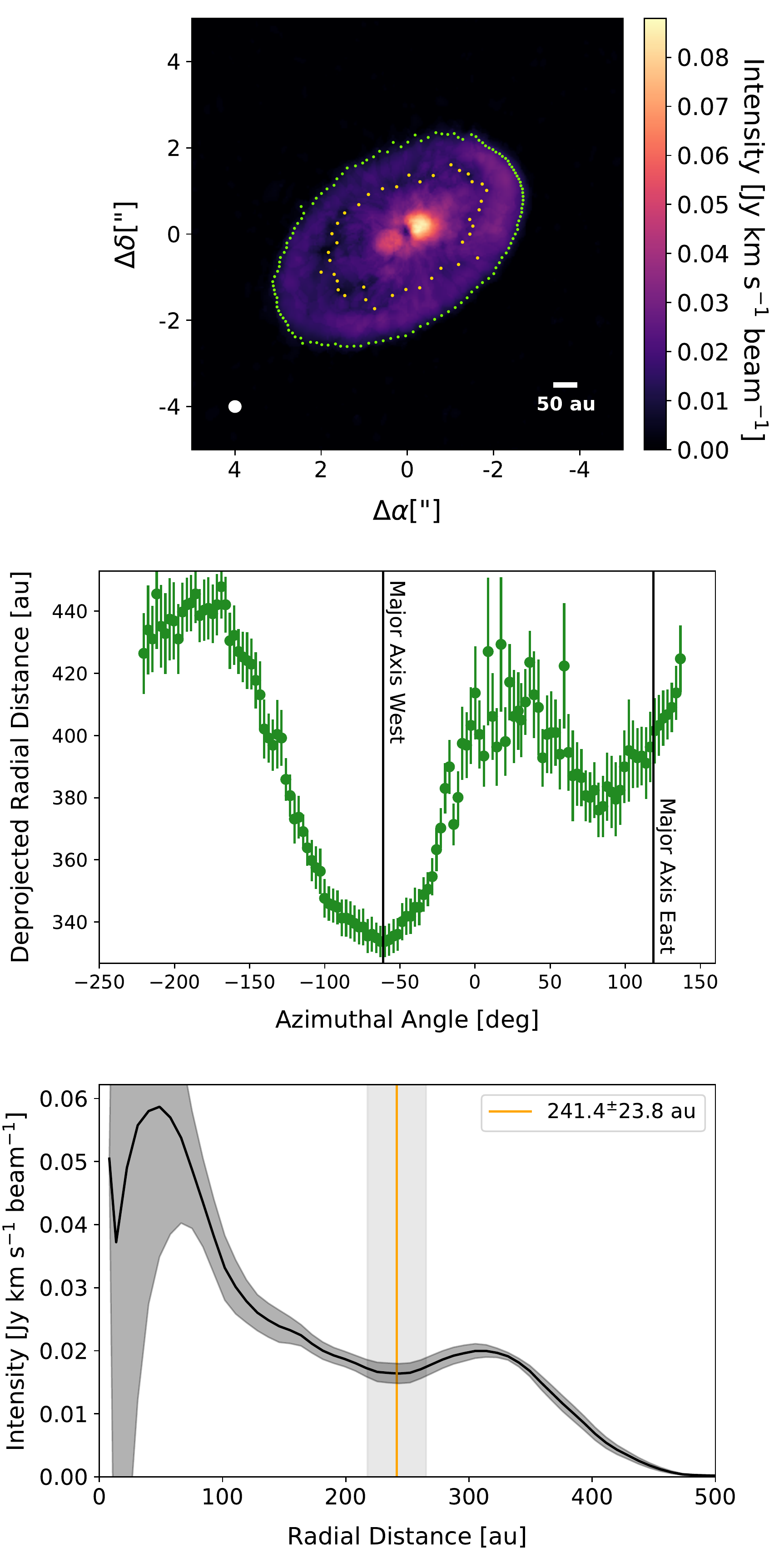}
      \caption{Top Panel: Integrated emission map of the C$^{18}$O gas emission. Yellow dots trace local minimum of emission, green circles trace the disk radial extent as the location at which 98$\%$ of the total azimuthal emission is included. Middle Panel: Radial distance from the center of the points tracing the border, azimuthal angle is measured from the North, to the East. Bottom Panel: Azimuthally averaged intensity profile of the C$^{18}$O emission, shaded area shows the 1$\sigma$ scatter at each radial bin divided by the number of beams spanning the angles over which the intensities are measured. Vertical yellow line marks the average radial location of the minimum of emission and the deviation of the data is indicated by the vertical grey region.
              }
         \label{C18O_gasgap}
\end{figure}

The channel maps for both $^{13}$CO and C$^{18}$O are shown in Appendix A. The known cloud absorption \citep[]{laura_elias, DSHARP_Huang_Spirals} affects the East side of the disk and blocks all $^{13}$CO emission from $v \ge 2.88$\,km s$^{-1}$, while for C$^{18}$O some cloud absorption is present near $v = 2.55-2.66$\,km s$^{-1}$. We note that the South part of the disk is the brightest. The systemic velocity is determined to be 1.95\,km s$^{-1}$ based on the C$^{18}$O channel map analysis using a spectral resolution of 0.05\,km s$^{-1}$ and kinematic modelling done with the \textsc{eddy} package \citep{2019JOSS....4.1220T}. 

Considering all spatial scales and imaging the whole FOV, extended, large-scale emission appears in both isotopologues around Elias 2-27 (see Figures in Appendix A). As $^{13}$CO is the most abundant isotopologue, the large-scale emission appears more strongly and along a wider range of velocities than in  C$^{18}$O. The extended emission is clearly identified between $v = 2.55-4.88$\,km s$^{-1}$ for $^{13}$CO, and between $v = 2.66-4.21$\,km s$^{-1}$ for C$^{18}$O. This large-scale emission appears to have a striping pattern, probably due to the lack of compact baselines in the observations. The shortest baseline is 15m, which projected in the sky for the observed frequencies recovers emission from $\lesssim$15$\arcsec$ scales, however, the detected emission may extend beyond our $\sim$20$\arcsec$ FOV. We compute integrated emission and mean velocity maps, shown in Figure \ref{mom_largescale}, from the channel maps that consider the whole FOV, including emission above 3$\sigma$. While in $^{13}$CO the large-scale emission is seen through the entire FOV, the C$^{18}$O large-scale emission is more constrained and crossing only through the East side of the disk. There is no clear velocity gradient between the large-scale emission and the disk in either isotopologue. 

From here on, we focus on the channel maps that trace the material closer to the disk rather than large-scale emission. These channel maps were obtained using uv-tapering and filtering of the short baselines as described in section 2. Figure \ref{mom0_size} shows the integrated emission (moment 0) of both CO molecules. Foreground absorption in $^{13}$CO is clear in the East side of the disk. Overlaying the continuum contours for the 0.89\,mm emission and measuring along the major axis, the East side of the disk appears to have a larger extent than the West side in the gas. Figure \ref{mom0_size} shows that the size difference is roughly 0.5” between East and West sides of the disk, in both CO tracers. At the source distance, this size difference corresponds to 58\,au between the projected emission extent of each side. To measure the East/West size variations, we trace the edge of the emission in the C$^{18}$O moment 0 map. The border of the disk is considered as the radius that encompasses 98$\%$ of the integrated emission at each sampled azimuthal angle (green points on the top panel of Figure \ref{C18O_gasgap}). We define the center based on the emission peak from the Band 7 continuum data and deproject accordingly, assuming the inclination and position angle  of the dust continuum. The deprojected radial distance as a function of azimuthal angle is shown in the middle panel of Figure \ref{C18O_gasgap}. Errors correspond to the astrometric error, calculated as described in section 3.1. The edge of the disk is not well described by a circle or an ellipse; it shows two local maxima and two local minima. The global minimum distance is located along the major axis on the West, but the global maximum distance is shifted with respect to the East major axis. The locations of maximum and minimum border extents are not colocated with the continuum spiral features, or their extension.

Another feature in the moment 0 map is the presence of a ``gap'' of emission at large disk radii in C$^{18}$O. To estimate its location, we trace the radial positions of emission minima sampling every 9$^{\circ}$, between 185-300\,au (this range is determined by analysis of the intensity profile). Using the mean value and standard deviation of the minima radial locations (yellow points in top panel of Figure \ref{C18O_gasgap}), we estimate the gap position at 241 $\pm$ 24\,au. In the $^{13}$CO integrated intensity map we do not observe a gap and cannot infer one from the intensity profile, even when excluding the azimuthal angles $\sim$65$^\circ$-155$^\circ$, where foreground absorption is strongest. If we trace the emission border of $^{13}$CO following the 98$\%$ integrated emission criteria, we obtain a similar emission border curve to the C$^{18}$O, shown in Appendix A.

\begin{figure*}[t]
   \centering
   \includegraphics[width=\hsize]{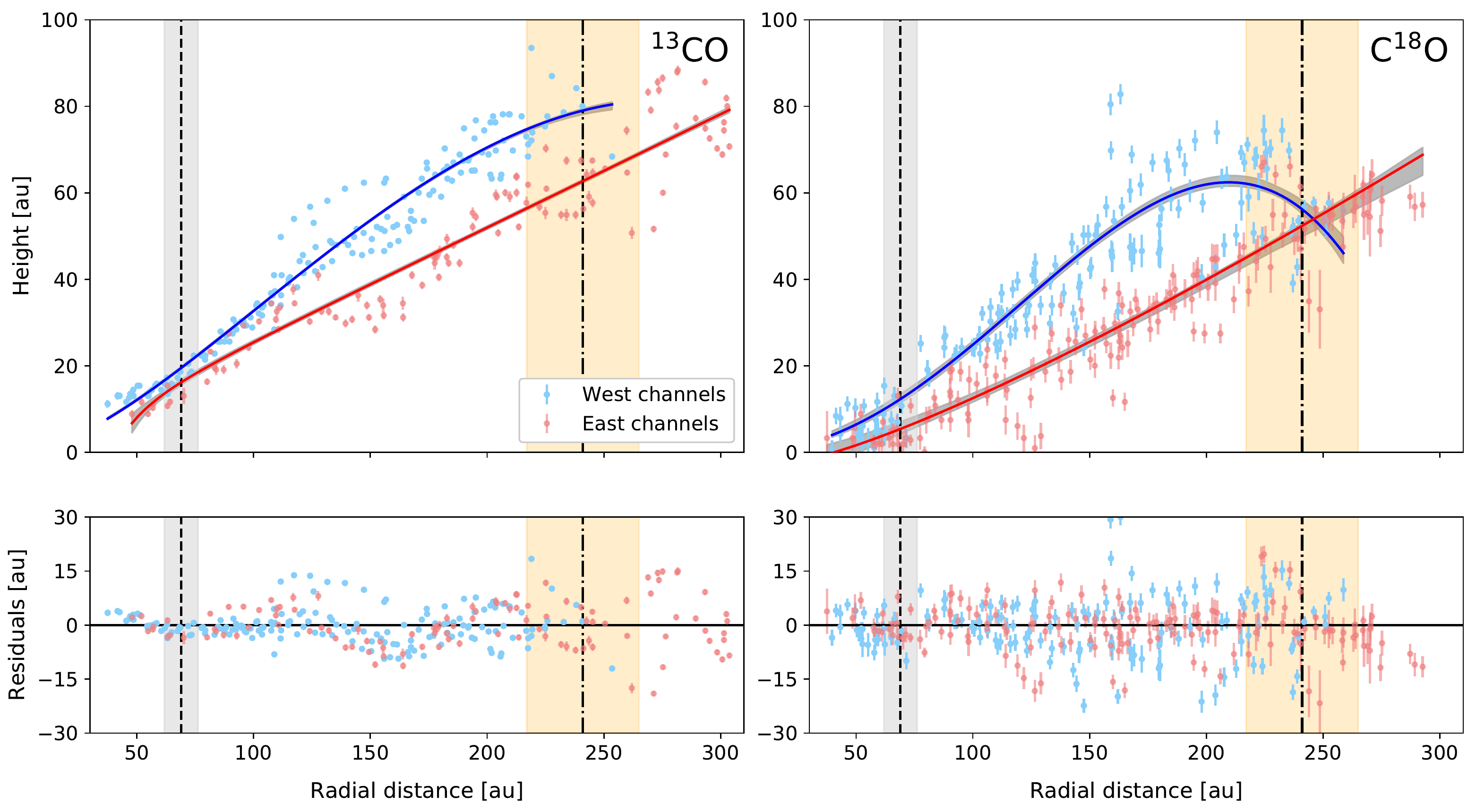}
      \caption{Top panels show the emission layer height as a function of radial distance to the star constrained from the $^{13}$CO (left) and C$^{18}$O (right) data. Blue points correspond to measurements coming from the West side of the disk, the red points come from the East side, colored line corresponds to the best-fit double power law height profile for the data, the grey area shows the uncertainty on the emitting layer as derived from the posteriors. The vertical dashed line indicates the location of the gap reported in the continuum, the dot-dashed line correspond to the gap location in the C$^{18}$O integrated intensity map. Grey area indicates the width of the dust gap \citep[obtained from][]{DSHARP_Huang_Annular}, orange area  indicates the gas gap's location uncertainty. Bottom panels show the residuals of each isotopologue after subtracting the best-fit model to the data.
              }
         \label{height_both}
\end{figure*}

\subsection{Tracing the emitting layer in $^{13}$CO and C$^{18}$O}

Using the method detailed in \cite{2018A&A...609A..47P} we recover the emission surface of each molecule. This is done by tracing the maxima from the upper layer of emission in the channel maps and applying geometrical relationships. The emitting layer is assumed to have a cone-like structure, this means that the height of the emitting layer should be symmetric with respect to the disk's major axis. 

If the variations in the projected radial extent of the disk found in Section 4.1 are related to variations in the emitting surface height then we would only expect symmetry along the disk's semi-major axis in the West side of the disk. The latter can be determined from Figure \ref{C18O_gasgap}, where we see that at the West Major Axis location ($\sim -60^{\circ}$) the radial distance of the emission border grows similarly when we move towards North (positive angles) or South (negative angles). This similarity in the radial distance increment maintains for $\sim 90^{\circ}$ in each direction (North and South), which corresponds to the complete West side of the disk. On the other hand, at the location of the East Major axis, we do not see this symmetry in the growth or decline of the radial distances towards North or South. The possible lack of symmetry is an important caveat and we expect this to affect mostly on the constraints obtained for the East side of the disk (where the variations in disk extension are larger). The emission surface we measure should be taken as a rough estimate. Details on the geometry relations can be found in the original publication \citep{2018A&A...609A..47P}.

From all available channels, we visually select those in which the top layer of gas emission can be clearly identified. For C$^{18}$O we select channels at velocities $+$0.77 to $+$1.55\,km s$^{-1}$ and $+$2.44 to $+$3.21\,km s$^{-1}$. For $^{13}$CO we select channels $+$0.66 to $+$1.66\,km s$^{-1}$ and $+$2.33 to $+$2.77\,km s$^{-1}$. 

The recovered height profile of the emission layer for both gas isotopologues is shown in Figure \ref{height_both}. Measurements obtained from channels that trace the East sides of the disk (with respect to the semi-minor axis) are colored red, while those from the West are blue. The $^{13}$CO and C$^{18}$O gas emission layer we constrain follows the expected distribution for these isotopologues in a disk: $^{13}$CO traces a higher layer from the mid-plane than C$^{18}$O in both East and West sides at all radii. The continuum and C$^{18}$O gas gap location are highlighted in Figure \ref{height_both}, together with the width for the dust gap \citep[as reported in ][]{DSHARP_Huang_Annular} and the uncertainty regions of each gap. No clear feature is recovered in the height profiles at the gap locations.

The height distribution is modelled using the equations for a complex flared surface presented in the \textsc{eddy} package \citep{2019JOSS....4.1220T}, such that the altitude of the emitting layer follows a double power-law:
\begin{equation}
    z(r) = z_0 \left(\frac{r}{r_0}\right) ^{\psi} + z_1 \left(\frac{r}{r_0}\right) ^{\varphi},
\end{equation}
where $r$ is the radial distance from the star, as measured in the azimuth plane, and the characteristic radius $r_0$ is fixed at 1$\arcsec$, corresponding to 116\,au for our system. The second power-law is aimed to be a correction on the first term, therefore we first find the best set of parameters for the emission layer characterized with a single power-law and then optimize close to those parameters, to include the second power-law. The best fit parameters of the model are assumed to be the median value from the posteriors of the MCMC simulations and are shown with their uncertainties (16th and 84th percentile uncertainties derived from the posteriors) in Table \ref{table_param_height}.

\begin{table}[h]
\def\arraystretch{1.5}
\setlength{\tabcolsep}{5pt}
\caption{Height model parameters from Channel analysis}
\label{table_param_height}      
\centering
\begin{tabular}{c c c c c}       
\hline\hline                
Param. & $^{13}$CO $-$ W &$^{13}$CO $-$ E & C$^{18}$O $-$ W & C$^{18}$O $-$ E \\    
\hline                       
            
   $z_0$ [au]& $77.6^{+8.2}_{-8.9}$ &   $30.0 \pm 0.2$ &       $44.0^{+7.3}_{-6.2}$ &      $32.3^{+8.9}_{-8.5}$ \\ 
   $\psi$    & $1.76^{+0.03}_{-0.04}$ & $1.01 \pm 0.01$ &       $2.19^{+0.12}_{-0.16}$ &    $1.14^{+0.11}_{-0.09}$ \\
   $z_1$ [au]& $-38.1^{+8.9}_{-8.1}$ &  $-0.3^{+0.1}_{-0.2} $ & $-11.8^{+6.1}_{-7.4}$ &     $-15.9^{+8.6}_{-8.9}$ \\ 
   $\varphi$ & $2.29^{+0.08}_{-0.04}$ & $-3.36^{+0.54}_{-0.52}$ &  $3.58^{+0.46}_{-0.24}$  &$0.46^{+0.15}_{-0.33}$ \\
\hline                                   
\end{tabular}
\end{table}

Our modelled emission surface presents consistent differences in the elevation and morphology between East and West sides of the disk. The height profiles have a quasi-linear form in the East channels, mostly tracing lower height values than the West. At larger radii ($>$200\,au) the West channels show a decrease in the emission surface height. Residuals obtained from subtracting the model from the observations are shown in the bottom panels of Figure \ref{height_both}. We see that the largest residual scatter is found between the dust and gas gap locations, which roughly coincides with the radial extent of the dust spiral arms (80-250\,au). The residuals from the $^{13}$CO emission show a ``curved'' pattern, indicating a more complex emitting surface. We note that in both isotopologues the residuals at the dust gap location are mostly negative, indicating a possible decrease in the emitting surface at this radii. This is unresolved with our spatial resolution. 

\subsection{Tracing the kinematics in $^{13}$CO and C$^{18}$O}

\begin{figure*}[t]
   \centering
   \includegraphics[width=\hsize]{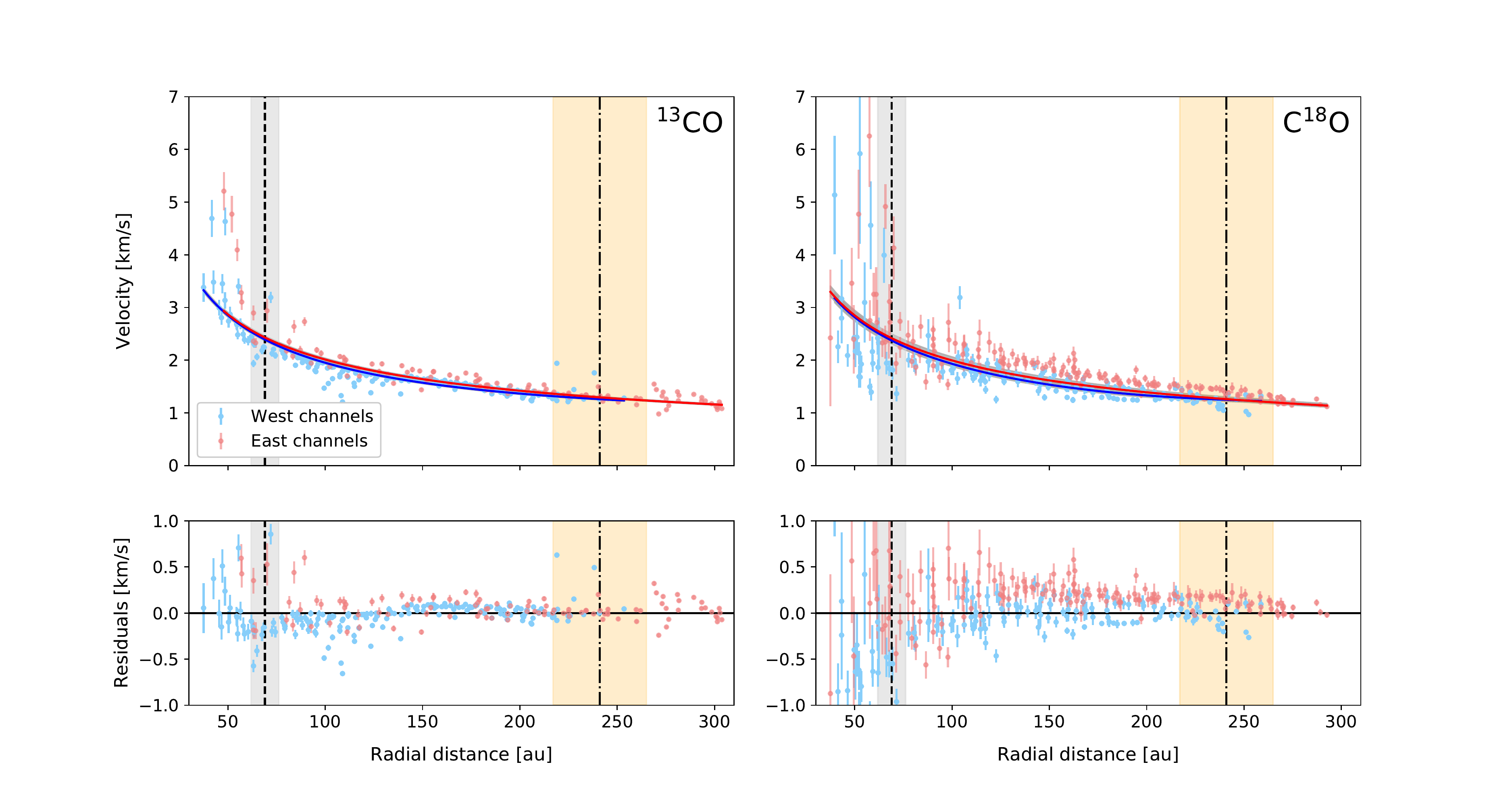}
      \caption{ Top panels show the data tracing the velocity of the gas emission, as a function of radial distance to the star, from the C$^{18}$O (left) and $^{13}$CO (right) isotopologues. Blue points correspond to measurements coming from the West side of the disk, red points come from the East side, plotted curves correspond to the best-fit Keplerian rotation profile and shaded area corresponds to the stellar mass uncertainty as indicated by the 16th and 84th percentiles of the posteriors. Vertical dashed line indicates the location of the gap reported in the continuum, dot-dashed line correspond to the gap location in the C$^{18}$O integrated intensity map. Grey area indicates the width of the dust gap \citep[obtained from][]{DSHARP_Huang_Annular}, orange area  indicates the gas gap's location uncertainty. Bottom panels show the residuals of each isotopologue after subtracting the best-fit model to the data.
              }
         \label{vel_both}
\end{figure*}

Besides tracing the height profile of the emitting layer, \citeauthor{2018A&A...609A..47P}'s method allows us to determine the velocity profile of the traced emission layer. In a given velocity channel, we know the projected radial velocity, $v_{obs}$, together with the systemic velocity for the source, $v_{syst}$. We are interested in determining the azimuthal velocity $v$ of a parcel of gas at an azimuthal radial distance $r$ and height $h$ from the star. Using the inclination angle $i$ and the polar azimuthal angle, $\theta$, we can relate the known velocities to $v$ through $v_{obs} = v_{syst} + v\cos(\theta)sin(i)$. To obtain $cos(\theta)$ we apply geometrical relationships from the measurements in the channel maps as defined in \cite{2018A&A...609A..47P}. The velocity profile will allow us to obtain a mass estimate for the central star and also test for super-Keplerian velocities at large radial distances, which is a characteristic expected in disks undergoing GI \citep{1999A&A...350..694B, 2007NCimR..30..293L}.

\begin{figure*}[t]
   \centering
   \includegraphics[width=\hsize]{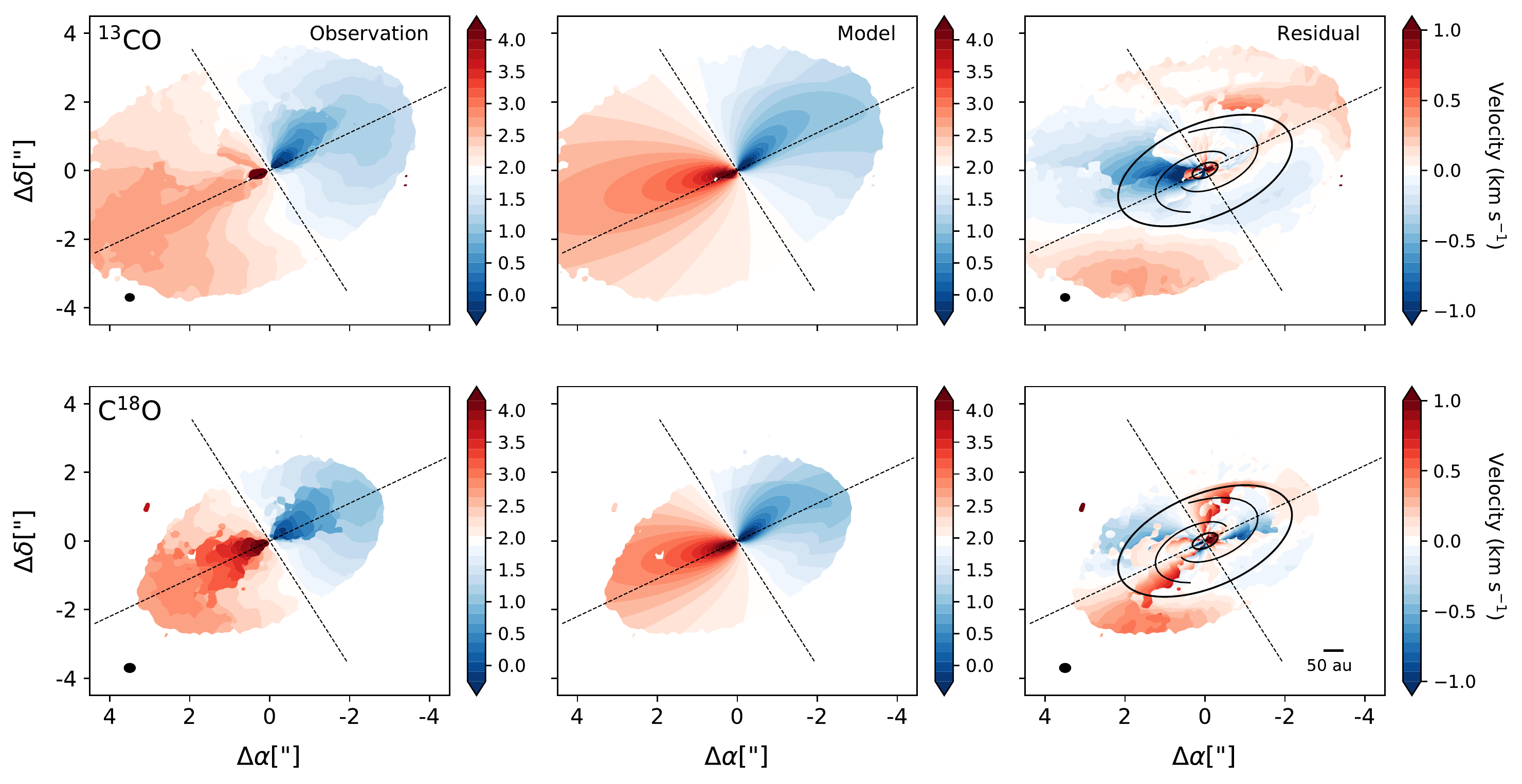}
      \caption{Velocity maps, model and residuals for $^{13}$CO (top row) and C$^{18}$O (bottom row). In each row, the first column shows the integrated emission velocity map (moment 1). The second column shows the velocity map model, computed using the constraints found for the emission surface and stellar mass. The third row shows the residuals calculated by subtracting the model map from the observations. The spiral arms show best fit parametric model from the 0.89\,mm dust emission, inner and outer ellipses indicate radial limits for the data used to derive the emission surface geometry and stellar mass values.
              }
         \label{kep_model_vel}
\end{figure*}

The velocity profiles of the emitting gas for both $^{13}$CO and C$^{18}$O are shown in Figure \ref{vel_both}. An overall velocity difference is observed between both sides of the disk, with measurements from the East side having a higher velocity. From our previous results (Figure \ref{height_both}) we know that the East side corresponds to the side apparently closest to the midplane. The difference in the velocity profile is in agreement with the height profile variations between sides, as being closer to the midplane results in larger velocities. We do not observe any noticeable behavior of the velocity profile at the location of the dust and gas gaps. 

We fit the velocity profiles with a Keplerian model to constrain the mass of the central star. Based on comparisons to stellar evolution models in the H-R diagram, the mass of Elias 2-27 has been reported to be $\sim$0.49M$_\odot$ \citep[]{2009ApJ...700.1502A, DSHARP_Andrews, laura_elias}. For our modelling, we incorporate the height distribution of each molecule, using the best fit double power-law model found previously (see Table \ref{table_param_height}). The modelled velocity at radial distance $r$ from the star will follow equation \ref{vel_prof}, where $G$ is the gravitational constant, $M_*$ is the central star mass, and $h$ is the height of the gas at azimuthal radius $r$:

\begin{equation}
    \frac{v^2}{r} = \frac{GM_* r}{(r^2 + h^2)^{\frac{3}{2}}}
     \label{vel_prof}
\end{equation}

We note that equation \ref{vel_prof} does not include the effects of the radial pressure gradient and the disk's self-gravity \citep{2013ApJ...774...16R}. 

We simultaneously fit the model to the data points from East and West sides of the disk, using MCMC simulations and taking into account the different height profiles (Figure \ref{height_both}). The curves for the expected Keplerian motion, considering the best-fit stellar mass and its 1$\sigma$ uncertainty range are shown over the data in Figure \ref{vel_both}. The final masses and errors are computed from the median value and 16th and 84th percentile uncertainties derived from the posteriors. From the $^{13}$CO measurements we constrain a stellar mass of $M_*=0.5 \pm 0.01$ M$_\odot$, while from the C$^{18}$O measurements we constrain $M_*=0.46^{+0.02}_{-0.03}$ M$_\odot$. Both values are in similar between them and compared to the previous estimates \citep[$M_* \sim 0.49M_\odot$,][]{2009ApJ...700.1502A, DSHARP_Andrews, laura_elias}.\\ 

Compared to the expected Keplerian velocity profile from the fits of Figure \ref{vel_both}, we see residuals throughout the whole radial extent. These simultaneous sub- and super-Keplerian velocities are expected if the emission layer height difference between the East and West sides was larger than what was constrained from the analysis of Figure \ref{height_both}. For now we only attempt to fit of a purely Keplerian rotation profile, but given the large disk mass of Elias 2-27 fitting a self-gravitating rotation curve is warranted. This possibility will be further explored in a separate paper (Veronesi et al., submitted).

Finally, we use the constrained emission surface of each side of the disk, and a stellar mass value of 0.49M$_\odot$, to build a model of the expected mean velocity maps of each isotopologue. The models are compared to the observations through residual analysis. Our model velocity maps consider only the Keplerian motion of the upper layer of the emission surface. In the case of Elias 2-27, the disk is inclined such that emission from the lower layer appears in the southern part of the disk \citep{DSHARP_Huang_Spirals}, which may be cause for larger residuals across the southern border. Additionally, our constraints on the shape of the emission surface do not cover the whole radial extent of emission, and are extracted from the data retrieved at radial distances between $\sim$40-300\,au ($\sim$0.35$\arcsec$ - 3.59$\arcsec$) from the central star (see Figure \ref{height_both}). Therefore, we should also expect to have larger residuals in the inner and outer regions where the emitting surface is not directly constrained by our method described above. 

Figure \ref{kep_model_vel} shows the $^{13}$CO and C$^{18}$O velocity maps in top and bottom rows, with observations, model, and residual velocity maps in left, middle, and right columns. The integrated velocity maps are computed using the \textsc{bettermoments} package \citep{2019ascl.soft01009T}, to accurately constrain the line of sight velocity from the channel maps with 0.111\,km s$^{-1}$ spectral resolution. Initial analysis of the observations allows us to identify marked perturbations throughout the disk, especially along the South and the West, where a distinct ``distorted'' pattern is observed in the outer disk. Along the major axis the C$^{18}$O data also displays ``distorted'' perturbations in a seemingly perpendicular form with respect to the azimuthal Southern ``distorted'' pattern. In the model maps we see that the West side of the disk is able to reproduce to some extent the ``distorted'' pattern along the South of the disk in both isotopologues, given the decrease in height in the outer disk. We do not see this pattern in the East side of the model map, as the retrieved emitting surface does not present large deviations from a linear cone-like model.

In the residuals, the radial extent that was used for determining the emission layer height profile is marked as two ellipses to define inner and outer radial bounds (40-275\,au). For larger radial distances than what was sampled, the residuals in the West side of the disk are much lower than the residuals in the East side, for both isotopologues. This means that the modelled West side emission layer with a ``dip'' in the emitting surface height at larger distances from the star is necessary when extending the model to larger radii. This indicates that the East side of the disk likely also has a decrease in the emitting surface height at larger radial distances. Given our limited range of sampled radial distances, we may not be sensitive to this ``turning point'' in the East. Within our sampled radii, marked by the ellipses, we observe negative residuals in the North-East quadrant of the $^{13}$CO emission. This roughly coincides with the location of the most prominent cloud absorption (see Figure \ref{mom0_size}), so we associate these residuals with the absorption. The C$^{18}$O residuals within the ellipses are much stronger and display a rough ``X'' shape across the center, with marked positive residuals close to minor and major axis' in North-West and South-East quadrants, respectively. As was noted, the radial distances within the ellipses coincide with the radial extension of the dust spiral arms, while the ``X'' shape does not have a clear co-location with the spiral structure, the largest positive residuals coincide with the location where a spiral starts and the other ends. This ``X'' shaped residual probably traces perturbations arising closer to the mid-plane, as it is not observed in $^{13}$CO and the C$^{18}$O emission traces a lower height layer.

\begin{figure*}[t]
   \centering
   \includegraphics[width=\hsize]{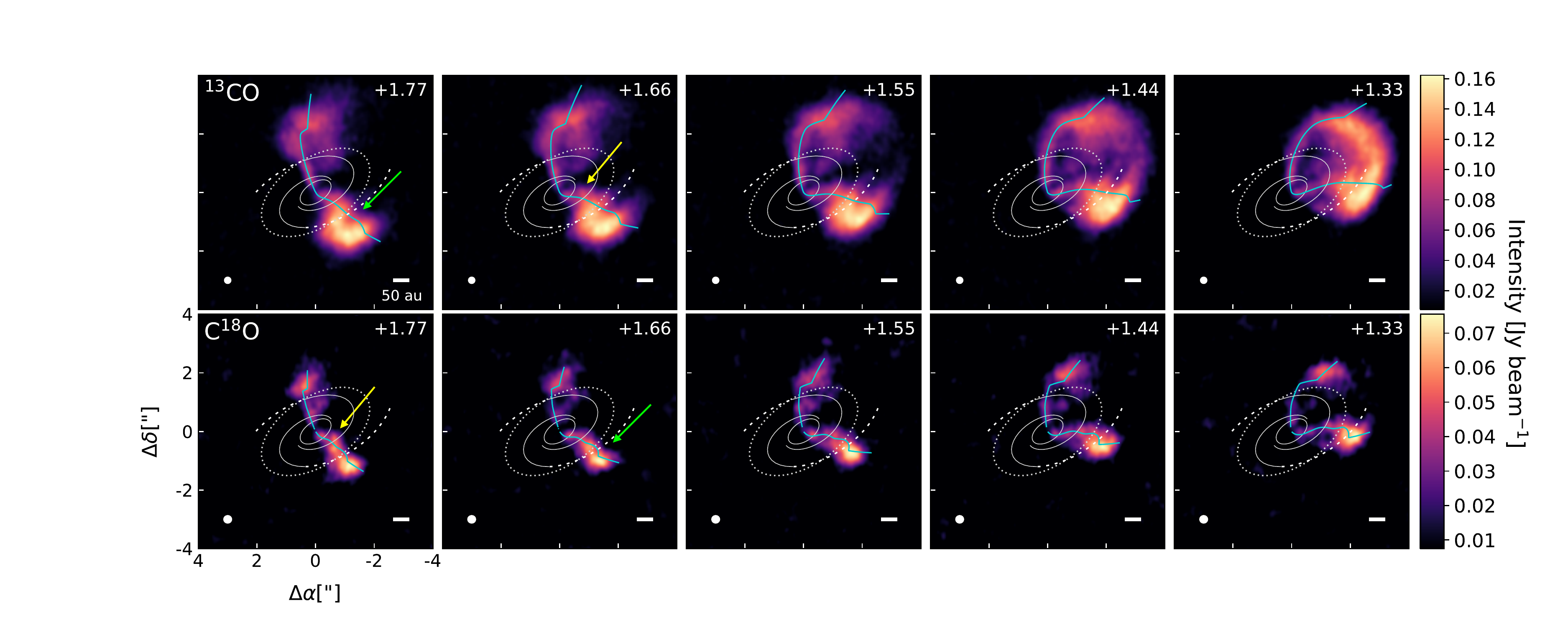}
      \caption{Selected central channels of $^{13}$CO (top) and C$^{18}$O emission (bottom). White continuous line shows the dust features: inner gap at 69\,au and the spirals as traced from the 0.89 continuum emission. Dotted white line traces the C$^{18}$O gas gap location at 241\,au. Dashed lines show how the spirals traced in the dust would extend further outside of the continuum emission. Blue curve traces the expected isovelocity curve of each channel, following the constrained emission layer geometry of the top layer. The velocity of each channel map is indicated in top-right corner of each panel, the beam size is in the bottom-left corner. Green arrows mark the outer perturbation, yellow arrows mark the inner perturbation.
              }
         \label{kinks_central}
\end{figure*}

\subsection{Features in the channel maps of $^{13}$CO and C$^{18}$O}

\begin{figure*}[t]
   \centering
   \includegraphics[width=\hsize]{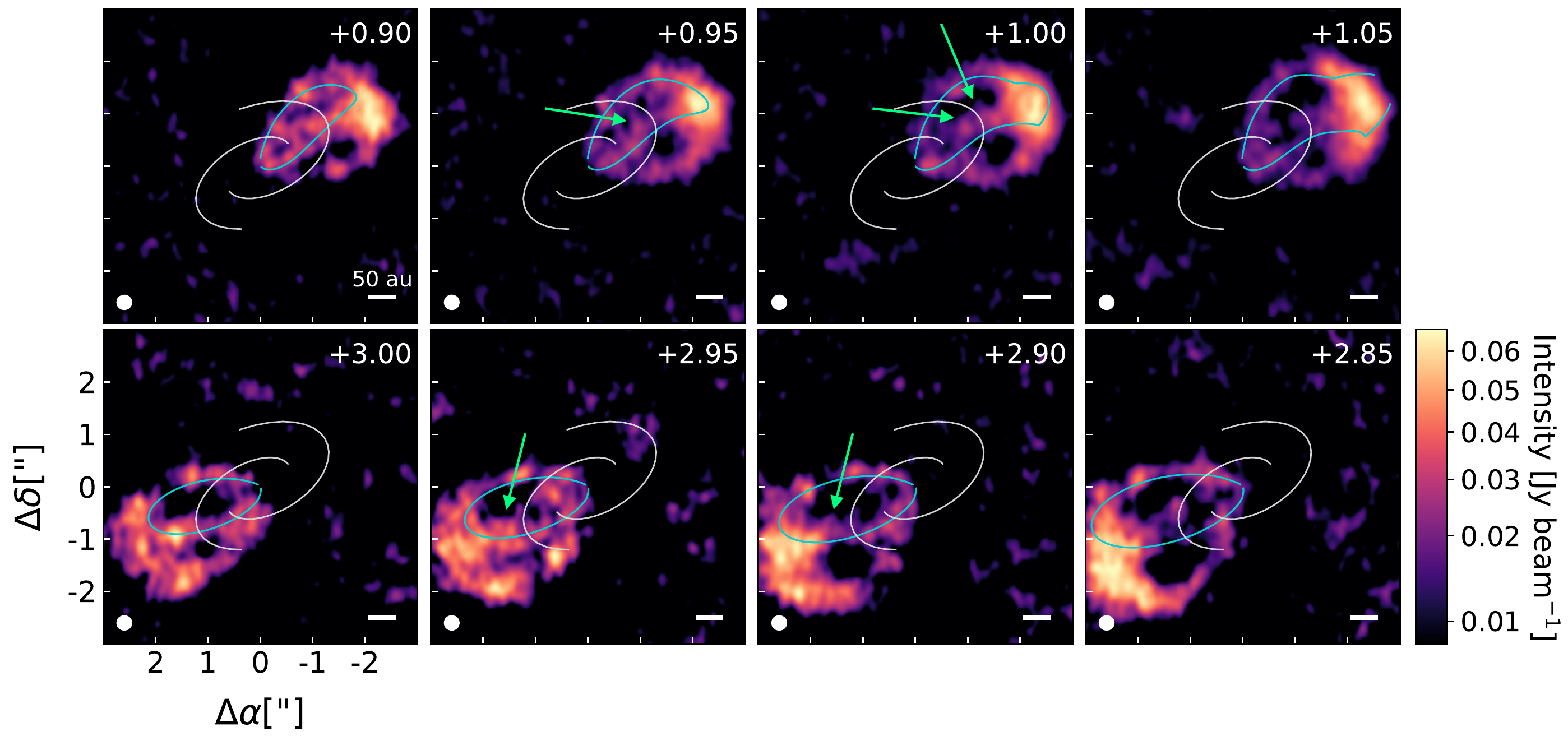}
      \caption{Selected high-velocity channels of C$^{18}$O emission. White lines trace the spirals detected in the 0.89\,mmcontinuum emission, blue lines indicate the isovelocity curves expected at each channel velocity, indicated in top right corner of each panel, following the constrained emission layer geometry of the top layer. Arrows indicate were deviations from expected isovelocity curves (``kinks'') are observed.
              }
         \label{kinks_small}
\end{figure*}

In the CO channel maps (see Figures in Appendix A) we observe several perturbations, which don't follow the expected Keplerian velocity field and we refer to them as ``kinks''. In the following figures, we overlay several previously characterized features of Elias 2-27 to use as reference: the continuum spirals and their extension, the location of the dust continuum and C$^{18}$O gaps, together with the expected isovelocity curves  for each channel. We note that the isovelocity curves seem perturbed as they are obtained from the model velocity emission shown in Figure \ref{kep_model_vel}, considering the constrained emission layer of each isotopologue and disk side. We only show the isovelocity curves of the top layer of emission, because the geometrical constraints have been derived for this layer and do not adequately trace the bottom layer, which we can visually identify. In the bottom layer, the emission seems to be coming from a layer at a further distance from the midplane than what we trace with the geometry of the top layer. This difference should be studied in future work and could be caused by the layers tracing different sectors of the disk due to temperature effects \citep[][]{2018A&A...609A..47P}, optical depth or some other asymmetry in the vertical disk structure. 

We observe two types of perturbations: inner kinks at roughly the location of the spiral arms (but outside the dust gap at $\sim70$ au), and outer kinks, beyond the extent of the continuum emission at $\sim 250$ au. Perturbations are strongly present in the central velocity channels of both CO tracers, shown in Figure \ref{kinks_central}. We observe the inner kink (marked with a yellow arrow, Figure \ref{kinks_central}) close to the spiral in the south side of the disk and along several channels, it appears strongest at channels +1.77 to +1.55\,km s$^{-1}$. This feature is co-located with the NW spiral. A large outer ``C'' shape (marked with a green arrow, Figure \ref{kinks_central}) can be seen beyond the gas gap in the south where the emission is brightest. This ``C'' shape feature is strongest in $^{13}$CO, along channels 1.55-2.33\,km s$^{-1}$, but in channels 1.66-1.88\,km s$^{-1}$ of C$^{18}$O we can also recognize it in the southern part of the disk. We suggest that the ``C'' is not a deviation from Keplerian motion, but rather is the projected emission from the upper and lower sides of the disk, connected by material bridging in the center between both sides. 

Besides these features located near the disk systemic velocity, we see more subtle deviations in high velocity channels, for both East and West sides of the disk in the C$^{18}$O channel maps. To highlight these perturbations, we show the expected isovelocity curves for these velocity channels, along with the dust spirals in Figure \ref{kinks_small}. The top and bottom panels of Figure \ref{kinks_small} show the West and East  C$^{18}$O emission, imaged with finer spectral resolution that is available only for the C$^{18}$O data. The deviations are most visible at 0.95-1.0\,km s$^{-1}$ in the West, where the top layer of the disk emission does not precisely follow the isovelocity curve (blue line) and appears perturbed at the spiral arm location. In the East side,  at 2.95-2.90\,km s$^{-1}$, similar deviations are apparent in the top emission layer of the disk, roughly co-located with the SE spiral. These kinks are not clearly observed in the $^{13}$CO maps, however it is expected that perturbations due to the spirals should be more apparent in  C$^{18}$O than in $^{13}$CO, as the C$^{18}$O traces a layer closer to the midplane, where the spirals reside. The deviations are better discerned in the West channels, possibly due to the lack of cloud absorption at these velocities, but also because, if the kinks are caused by the spiral arms, the highest-contrast spiral is in the West side of the disk. 

Recently, \cite{2020ApJ...890L...9P} reported the presence of a kink in the Northern side of Elias 2-27, at the location of the dust gap, which was signaled as possible indicator of a planetary companion. We do not recover this feature, possibly given our lower spatial resolution: the DSHARP data has 4-5 times better angular resolution than this work, and is sensitive to spatial scales down to $\sim$6\,au in this system. This work analyses data with higher spectral resolution (3-6 times better) and less affected by cloud absorption than previously published studies. The latter makes us sensitive to the perturbations reported in this work, however we are not able to detect small spatial scale perturbations due to our angular resolution.

\section{Hydrodynamic Simulations of a Disk undergoing GI}

Elias 2-27 has been subject to different modelling approaches in order to explain the origin of the observed spiral substructure \citep[]{2017ApJ...835L..11T, 2017ApJ...839L..24M, 2018MNRAS.477.1004H, 2018ApJ...860L...5F, 2018ApJ...859..119B, 2020MNRAS.498.4256C}, with most of the modelling efforts oriented towards a gravitationally unstable disk as the disk-to-star mass ratio ($q$) has been estimated to have values around 0.2-0.3 \citep[]{2009ApJ...700.1502A, 2009ApJ...701..260I, 2010A&A...521A..66R, laura_elias} and gravitational instabilities are expected  when disk-to-star mass ratios are $>0.1$ \citep{2016ARA&A..54..271K}. The amount of spiral arms (m) excited in a GI scenario will depend inversely on the disk-to-star mass ratio (m $\sim M_*/M_d$). Given the m$=$2 spiral mode observed in Elias 2-27, previous simulations \citep{2017ApJ...835L..11T, 2017ApJ...839L..24M, 2018MNRAS.477.1004H, 2018ApJ...860L...5F} have aimed at producing the system using higher disk mass estimates ($M_d/M_* \sim0.5$) than those derived from the observations \citep[$M_d/M_* \sim0.2-0.3$, ][]{laura_elias}. Recent work by \citeauthor{2020MNRAS.498.4256C}, however, has shown that a disk-to-star mass ratio of 0.27 may also reproduce the observations. While this low disk-to-star mass ratio predicts a high amount of spirals, it has been shown that ALMA sampling can make a disk of $q=$0.25 appear as a $m=$2 system \citep{2014MNRAS.444.1919D}.

We performed a total of 10 three-dimensional, dusty, gaseous hydrodynamical simulations using the SPH code \texttt{PHANTOM} \citep{2018PASA...35...31P}. To accurately compare the simulations to our multiwavelength observations, we use the multigrain setup considering 5 different grain sizes, ranging from 1\,micron to 1\,cm in 5 logarithmically-spaced size bins, assuming a size distribution of $dn/da\propto a^{-3.5}$.  Multiple grain sizes are necessary as, while the most efficient emission at wavelength $\lambda$ comes from dust grains of size $a\sim\lambda/2\pi$ \cite[e.g.][]{2006ApJ...636.1114D}, there is an overall contribution from all grains. The dust is modelled self-consistently with the gas, using the multigrain ``one-fluid'' approach, where we limit dust flux using the Ballabio switch \citep{2018MNRAS.477.2766B}. Since the disk is massive and self-gravitating, the dust remains in the strongly-coupled regime ($St\lesssim 1$) out to grain sizes of several cm, so we do not need to use the two-fluid approach. In this regime, the dust exerts a force back on the gas (back-reaction) that is significant \citep{2018MNRAS.479.4187D}, so we include this effect on the gas.

In all 10 simulations we used 1 million SPH particles and assumed a central stellar mass of 0.5M$_{\odot}$, represented by a sink particle  \citep{1995MNRAS.277..362B}, with accretion radius set to 1\,au. We set the initial inner and outer disk radii to 5\,au and 300\,au respectively. Different simulations vary in disk-to-star mass ratio and density profile. The total dust mass in the system is kept constant at $0.001$ M$_{\odot}$, since this is observationally constrained \citep{laura_elias}. As the total dust mass is a fixed value in our simulations, to obtain different disk-to-star mass ratios we vary the gas-to-dust ratio ($\epsilon$), which corresponds to $\epsilon=100,151$ and 252 for $q=0.2, 0.3$ and 0.5, respectively. We do not sample lower gas-to-dust ratios because we do not expect to recover a m = 2 spiral arm morphology for $q<$0.2. While sampling the emission with ALMA can make disks with high $m$ values appear as $m=$2 systems, when going to very low $q$ this effect does not hold \citep{2014MNRAS.444.1919D}. The sound speed profile was set as $c_\mathrm{s}\propto R^{-0.25}$, and we used two surface density profiles: either a simple power-law,

\begin{figure*}[t]
   \centering
   \includegraphics[width=\hsize]{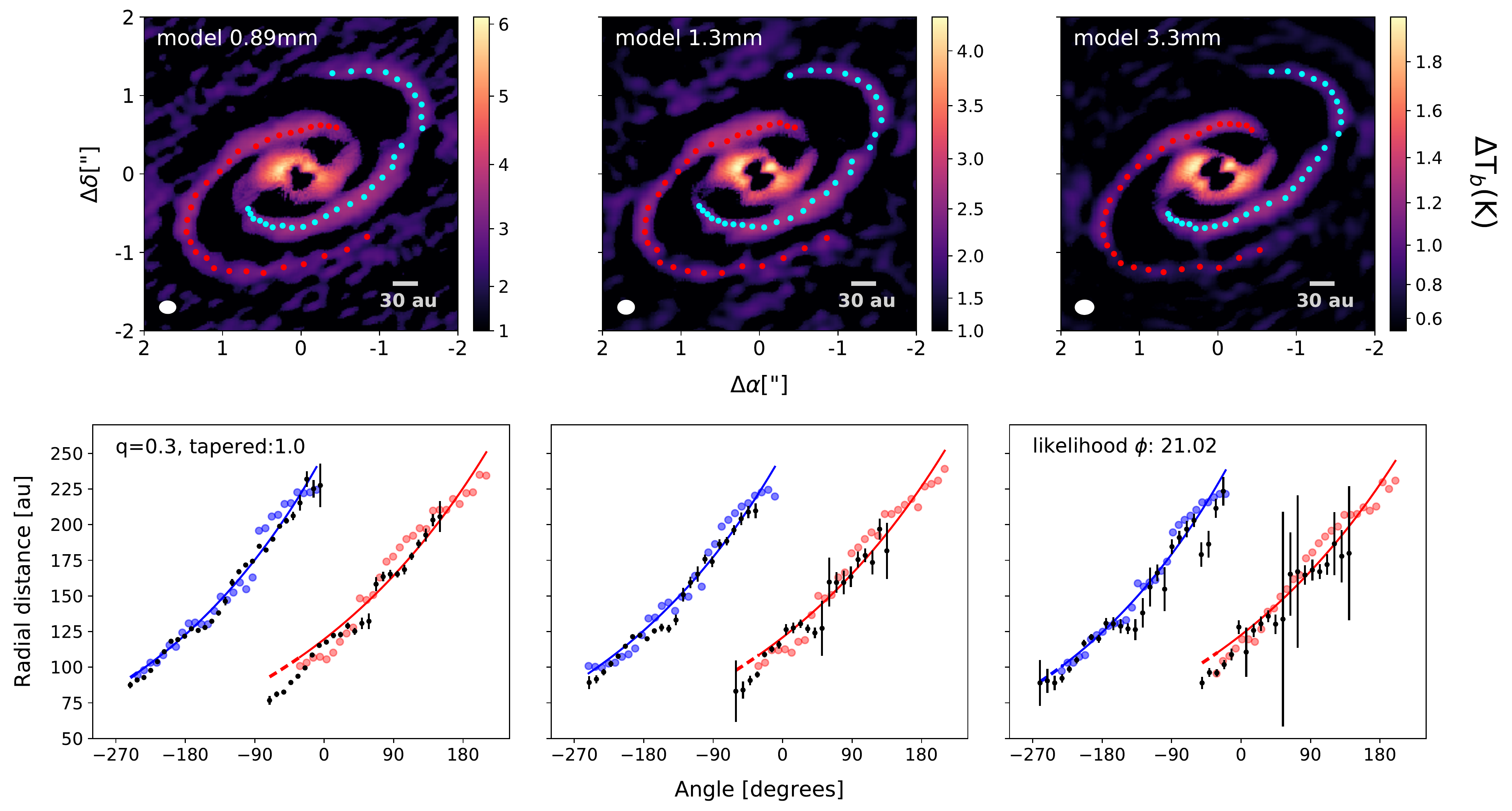}
      \caption{Panels from Left to Right correspond to data from the 0.87\,mm, 1.3\,mm and 3\,mm simulated observations for a exponentially tapered dust density profile with index 1.0 and disk-to-star mass ratio $q$=0.3. Top: images of simulated emission with subtraction of azimuthally averaged intensity profile, blue and red dots trace the maxima location along the spirals. Bottom: blue and red dots correspond to the deprojected radial location of the traced spirals in the simulated emission. Colored solid lines show the constant pitch angle logarithmic spiral fit, dashed colored lines extend the fit to lower radii. Black points are the deprojected radial location of the spirals from the observations (Section 3) and their astrometric error. Pitch angle likelihood parameter is indicated in the bottom right panel.
              }
         \label{model_q03_98}
\end{figure*}

\begin{equation}
\label{eq_pl} 
    \Sigma(R) = \Sigma_{0}\left(\frac{R}{R_{0}}\right)^{-p}
\end{equation}

\noindent
where $\Sigma_0$ is the surface density at the inner edge of the disk, and $p$ either 1.3 or 1.5, or an exponentially-tapered power-law,

\begin{equation}
\label{eq_pl_exp} 
    \Sigma(R)=\Sigma_{c}\left(\frac{R}{R_{0}}\right)^{-p} \exp \left[-\left(\frac{R}{R_{c}}\right)^{2-p}\right]
\end{equation}

\noindent
where $R_c$ is the characteristic radius of the profile, which we set to $R_c=200$au and $p$ either 0.7 or 1.0. In both surface density profiles $R_0$ is the reference radius and is set to $R_0=10$au. 

We used a polytropic equation of state, and assumed that the disk cooled through the $\beta$ cooling prescription \citep{2001ApJ...553..174G}, where the cooling timescale, $t_\mathrm{c}$, is related to the dynamical timescale, such that $t_\mathrm{c} = \beta t_\mathrm{dyn}$. The dynamical timescale is the rotation period, $2\pi/\Omega$, and we set $\beta = 15$. Finally, each simulation is computed for 10 orbital periods at the outer radius (300\,au), from which we receive outputs every 0.1 fraction of an orbital period. The detail of the model parameters are shown in Table \ref{table_param_model}.

\begin{table}[h]
\def\arraystretch{1.5}
\caption{SPH Model Parameters}
\label{table_param_model}      
\centering                          
\begin{tabular}{c c}       
\hline\hline                
Param. & Value\\    
\hline                       
            
    $M_*$ [$M_\odot$]  & 0.5\\ 
    $R_{in}$ [au]     & 5\\
    $R_{out}$ [au] & 300\\
    $M_{dust}$ [$M_\odot$] & 0.001\\
    gas-to-dust mass ratio & 100, 151, 252\\
    Min. Grain size [cm]& 10$^{-4}$ \\
    Max. Grain size [cm]& 1\\

\hline                                   
\end{tabular}
\end{table}

While SPH simulations portray the overall dynamic behaviour of a system, in order to accurately compare the model to an observation it is necessary to produce radiative transfer calculations of the SPH outputs, and then simulate mock observations using the same observing conditions (uv coverage) as the actual observations. Radiative transfer is necessary because it accounts for the multiwavelength emission the grains will have given the stellar characteristics and grain distribution. Sampling the radiative transfer output with the same observing configuration is crucial, for the antenna distribution and observation time will result in a given angular resolution and will form an image according to the sampled uv-coverage.

\subsection{Dust Simulations}

\begin{figure*}
   \centering
   \includegraphics[width=\hsize]{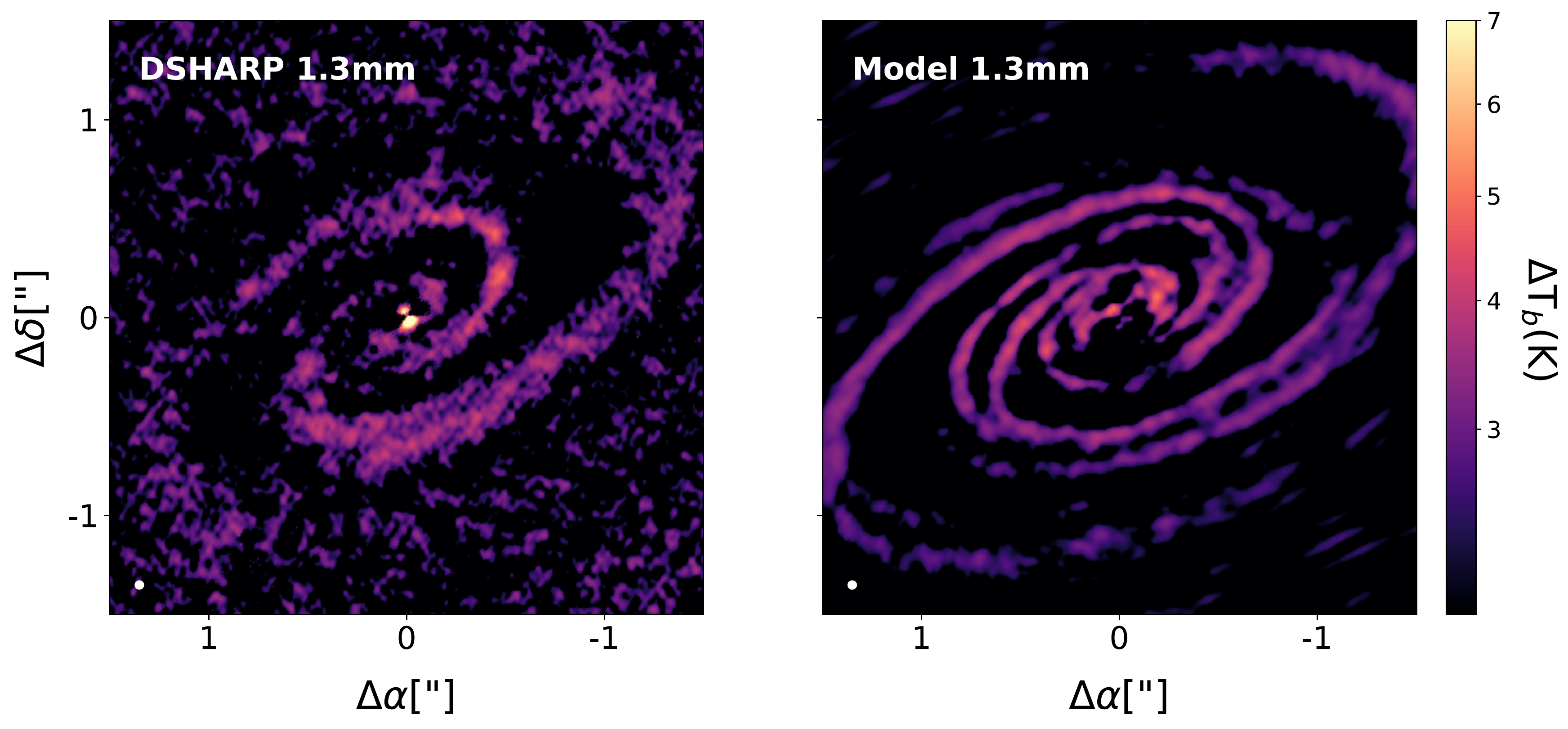}
      \caption{Residual emission after subtracting the azimuthally averaged intensity profile. Left: DSHARP \citep{DSHARP_Andrews} high-angular resolution emission at 1.3\,mm. Right: Simulated emission of a GI disk with an exponentially tapered dust density profile of index 1.0 and disk-to-star mass ratio $q=$0.3. Both data sets have the same angular resolution, indicated by the beam in the bottom left of each panel.
              }
         \label{DSHARP_comp}
\end{figure*} 

The Monte Carlo radiative transfer \textsc{mcfost} code \citep{mcfost1,mcfost2}  was used to compute the disk thermal structure and synthetic continuum emission maps at each observed wavelength (0.89\,mm, 1.3\,mm and 3.3\,mm). We assumed $T_\mathrm{gas} = T_\mathrm{dust}$, and used $10^7$ photon packets to calculate $T_\mathrm{dust}$. We set the parameters for the central star to match those of the Elias 2-27 system \citep{2009ApJ...700.1502A, laura_elias}, with temperature $T=3850$ K, $M = 0.5$M$_{\odot}$ and $R_* = 2.3$ R$_\odot$. 

To create the density structure as input into the \textsc{mcfost} calculation, each SPH simulation underwent Voronoi tesselation such that each SPH particle corresponds to one \textsc{mcfost} cell. We assumed the dust is a mixture of  silicate and amorphous carbon \citep{1984ApJ...285...89D} and optical properties were calculated using Mie theory. The grain population consists of 100 logarithmic bins ranging in size from  0.03 $\mu$m to 1\,mm. The dust density of a grain size $a_i$ was obtained by interpolating from the SPH dust sizes in each cell in the model. We assume that grains smaller than half the smallest SPH grain size (0.5 $\mu$m) are perfectly coupled to the gas distribution. We normalised the dust size distribution by integrating over all grain sizes, where a power-law relation between grain size $a$  and number density of dust grains $n(a)$ was assumed such that d$n(a)\propto a^{-3.5}$\,d$a$.

The radiative transfer emission map of each wavelength was sampled with the same uv-coverage as the observations at each corresponding wavelength, using \textsc{galario} \citep{2018MNRAS.476.4527T} to create mock ALMA visibilities that were afterwards processed using the same deconvolution procedures as with the observations (described in Section 2) to obtain the final mock ALMA images.

For each simulation image we subtract its azimuthally-averaged radial profile of emission, following the same procedure described to trace the spiral morphology on the multi-wavelength  observations (Section 3.1). We find that most of our models are able to accurately reproduce the m$=$2 large-scale spiral morphology. This is expected in the ALMA images, even at lower disk-to-star mass ratios (were we expect larger number of spiral arms), as was shown by \cite{2014MNRAS.444.1919D}. To select the simulation that best resembles our observations, we measure the spiral's pitch angle and compare them to the observational values. The pitch angles measured for each spiral and model setup are shown in Table \ref{table_pitch_model}. The model pitch angle ($\phi_{model}$) and the observational pitch angle ($\phi_{obs}$) are compared considering the difference between the values of each spiral and wavelength and weighing by the error of the observational pitch angle ($\sigma_{obs}$), following a likelihood parameter determined by:

\begin{equation}
    \sum_{\lambda} \sqrt{\left[\left(\frac{\phi_{model}^{NW} - \phi_{obs}^{NW}}{\sigma_{obsNW}} \right)^2 + \left(\frac{\phi_{model}^{SE} - \phi_{obs}^{SE}}{\sigma_{obsSE}}\right)^2 \right]_{\lambda}}
\end{equation}

The simulation that best reproduces the observations follows a dust density profile of an exponentially tapered power-law with index 1.0 and a disk-to-star mass ratio of 0.3. These parameters are close to the previously published observational constraints for this disk \citep{laura_elias} and similar to the result derived in \citeauthor{2020MNRAS.498.4256C}. The simulation is shown in Figure \ref{model_q03_98}. Besides reproducing the pitch angle values, the radial extension of the spirals and their overall morphology in the simulation is similar to the observations. We note that the comparison to each simulation set was made for a specific timestep within all the simulation outputs, and the selection was based on the output that showed a clear 2 spiral arm feature after at least 6 outer orbits. For the case of the best-fit simulation, the selected timestep was after 6.4 outer orbits (at 300\,au from the star). It is important to state that, as expected, the spirals arising from GI are constantly excited and de-excited throughout the timelapse of our simulation. This means that for a same set of parameters, there may be several timesteps that accurately reproduce the morphology but others where no spirals are seen.

Even though there is a simulation that reproduces the observations better than the rest, the pitch angles of the different SPH simulations are in most cases comparable to those of the observations ($\sim$ 12.9$^{\circ}$ and $\sim$ 13.2$^{\circ}$, for NW and SE spiral respectively, with small variations between wavelengths). This shows that it is possible to reproduce the grand-design spirals even at lower disk-to-star mass ratios than previously tested in this system \citep[]{2018MNRAS.477.1004H, 2017ApJ...839L..24M}, with stellar mass, disk dust mass and density profile values comparable to the observational constraints. The likelihood value of the rejected models is shown in Table \ref{table_pitch_model}. Additionally, we see that not all GI spirals, recovered from the models and measured with our method (section 3.1), are perfectly symmetric (same pitch angle), which is a property that has been predicted in other works for GI excited spiral arms \citep{2018ApJ...860L...5F}. This is specifically observed in the models with q=0.2. 

We compare our best-fit model with the high-angular resolution DSHARP \citep{DSHARP_Andrews} data, at 1.3\,mm. The comparison of the subtracted, residual images is shown in Figure \ref{DSHARP_comp}. From the visual comparison we clearly see that the internal structure of the spirals are different. Even at high-angular resolution, the observations show thicker and wider, continuous spirals. On the other hand, the simulation shows thinner and discontinuous spirals, which are made from a superposition of filaments. At larger radii we note that the spirals of the observation get wider, while in the simulation they remain thin. The causes for these differences are  discussed in section 6.3.

We measure the pitch angle value, sampling the same angular extent as in the work by \cite{DSHARP_Huang_Spirals}. We retrieve a value of 15.56$^{\circ} \pm 0.06^{\circ}$ in the NW spiral and 12.76$^{\circ} \pm 0.06^{\circ}$ in the SE spiral. These values differ from the constraint from the lower angular resolution simulations, showing that the beam smearing does impact in the pitch angle measurements, as proposed in section 3.1. The pitch angle values for the DSHARP observations are 15.7$^{\circ} \pm 0.2^{\circ}$ in the NW spiral and 16.4$^{\circ} \pm 0.2^{\circ}$ in the SE spiral. The main difference is from the SE spiral, possibly related to the effects of the morphology differences and the thickness of the spiral from the observation.

\subsection{Gas Simulations}

\begin{figure}
   \centering
   \includegraphics[width=\hsize]{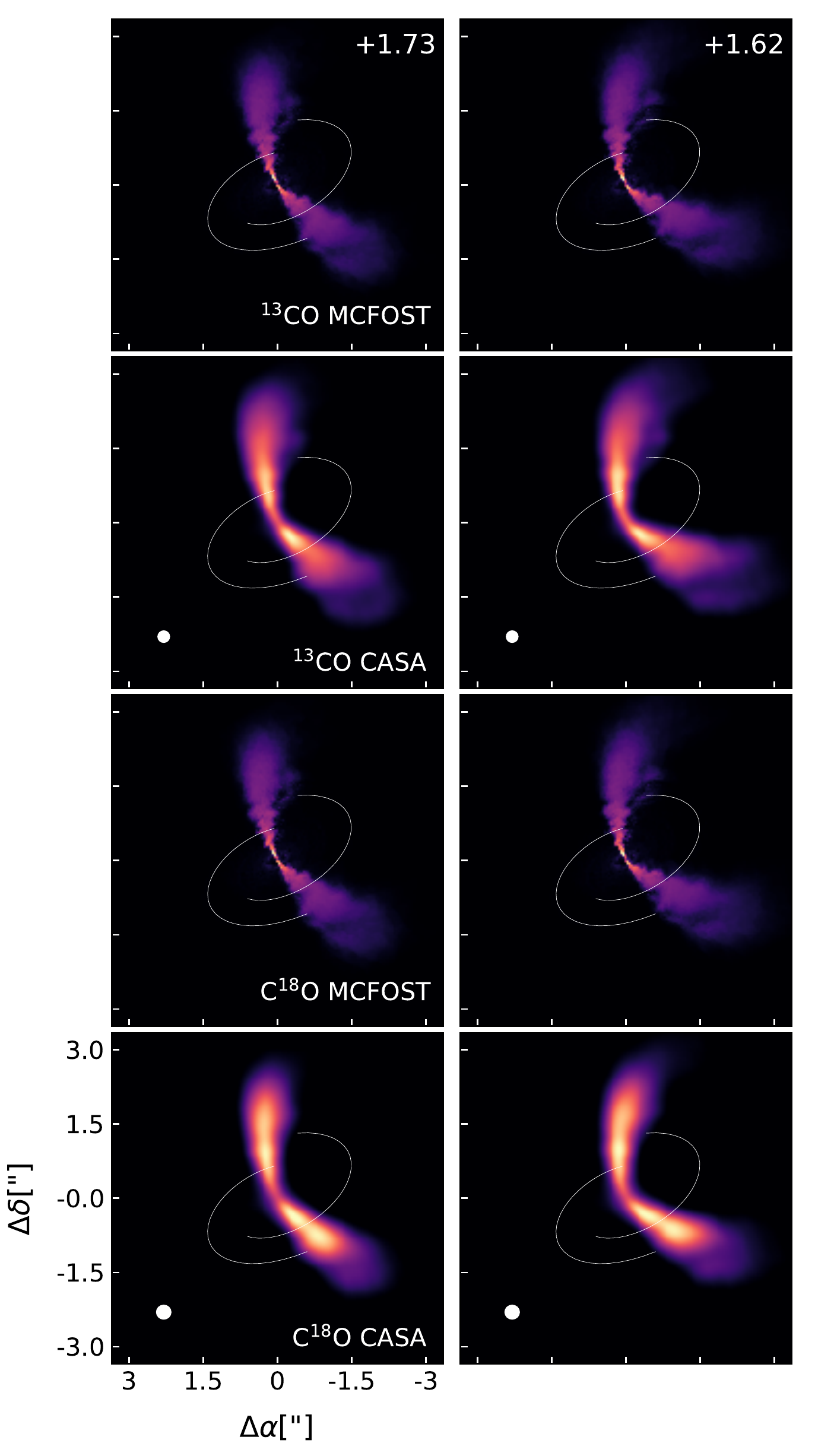}
      \caption{Individual emission of channels at velocities +1.73km/s (Left) and +1.62km/s (Right) for simulated $^{13}$CO (Top two rows) and C$^{18}$O (Bottom two rows) emission. From top to bottom, the first and third rows correspond to \textsc{mcfost} output emission. The second and fourth rows shows the emission after applying uv-coverage as in observations and processing with CASA. White lines trace the spirals from the best simulation (q=0.3, exponentially tapered dust density profile index 1.0). The beam for the simulated ALMA images is shown in the bottom left of each corresponding panel.
              }
         \label{model_gas_all}
\end{figure} 

\begin{figure*}
   \centering
   \includegraphics[width=\hsize]{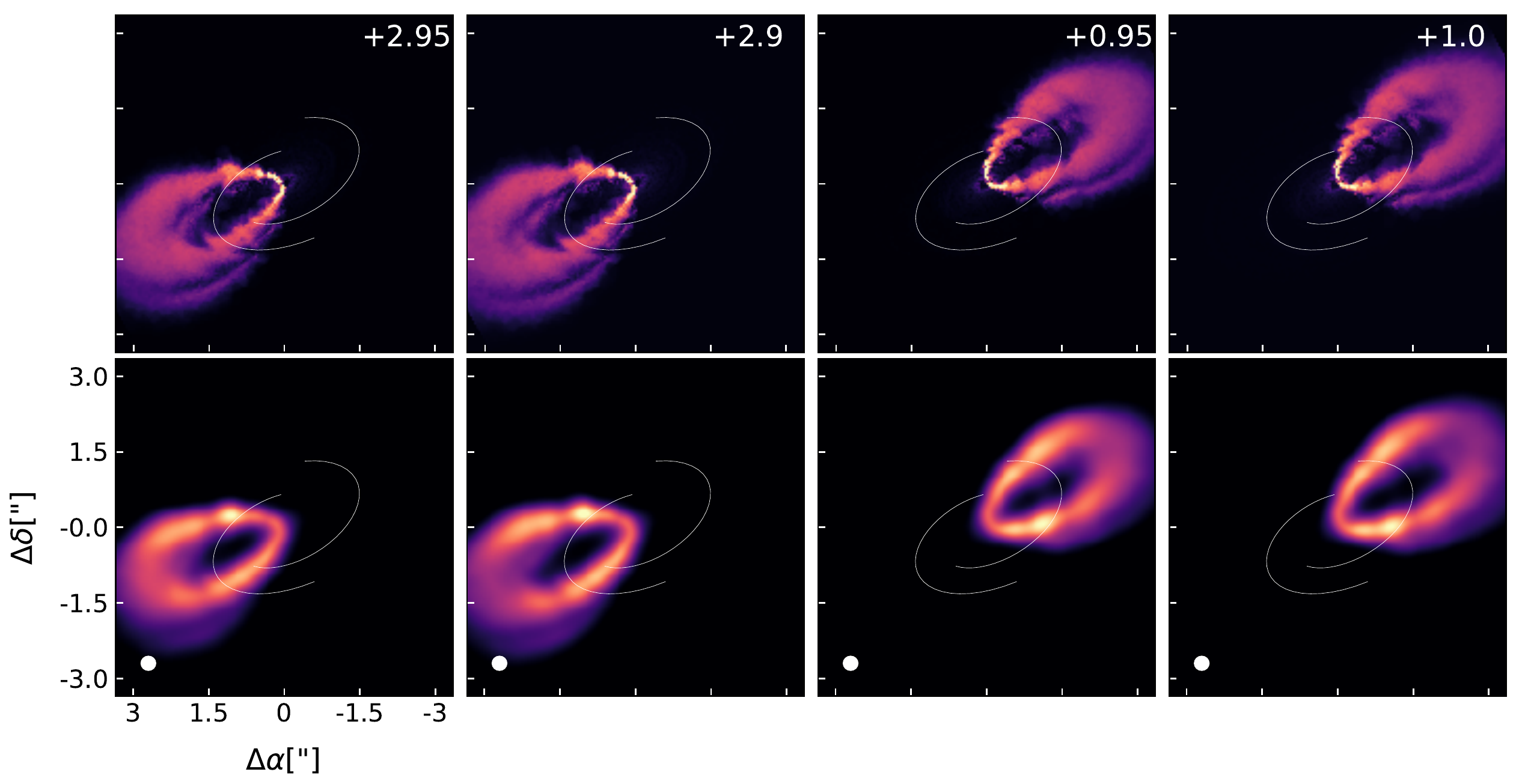}
      \caption{Individual emission of selected channels (corresponding velocities marked in top row) for C$^{18}$O simulated emission. Top row corresponds to \textsc{mcfost} output emission. Bottom row shows the emission after applying uv-coverage as in observations and processing with CASA. White lines trace the spirals from the best simulation (q=0.3, exponentially tapered dust density profile index 1.0). The beam for the simulated ALMA images is shown in the bottom left of each corresponding panel.
              }
         \label{model_gas_small18}
\end{figure*} 

From the simulation that best reproduces the pitch angle of the spiral arms (q = 0.3, density profile with a tapered power-law of index 1.0), we compute the simulated channel maps. As with the dust simulations, we use \textsc{mcfost} \citep{mcfost1,mcfost2}, with the same parameters as before, to compute the disk thermal structure and synthetic $^{13}$CO $J = 3 - 2$ and C$^{18}$O $J = 3 - 2$ line maps. The molecule abundances, relative to local H$_2$ are set to 7 $\times$ 10$^{-7}$ for $^{13}$CO \citep[as done in ][]{2020arXiv200715686H}  and 2 $\times$ 10$^{-7}$ for C$^{18}$O  \citep[following the estimate from][]{1982ApJ...262..590F}. The spectral resolution for the kinematic simulations is set to 0.111\,km/s to match the observations and we additionally compute C$^{18}$O channel maps with 0.05\,km/s resolution, to compare with the finer spectral resolution data. 

The results for two representative channel maps are shown in Figure \ref{model_gas_all}. While in the radiative output we observe the characteristic  ``GI-wiggle'' shown by \cite{2020arXiv200715686H} at the spiral arm's location, when sampling the data with the observation's uv-coverage the  ``GI-wiggle'' is not visible in any isotopologue. In the work by \cite{2020arXiv200715686H} the  ``GI-wiggle'' remains visible even after convolving with a gaussian beam. Compared to \cite{2020arXiv200715686H}, our simulated ALMA images  have a $\sim$4 times lower spectral resolution (0.111\,km/s compared with 0.03\,km/s) and  $\sim$3 times lower angular resolution (0.3$\arcsec$ compared to 0.1$\arcsec$). While the higher spectral resolution C$^{18}$O observations have a comparable channel width (0.05\,km/s) with the analysis of \cite{2020arXiv200715686H}, the angular resolution smears the  ``GI-wiggle'' features (see Figure \ref{model_gas_small18}). We note that no North/South brightness asymmetry is present in the simulated channel maps. Additional channel maps computed with a 90$^\circ$ inclination do not show any significant vertical difference between East/West sides.

\section{Discussion}

\subsection{Spiral structure and multi-wavelength dust continuum emission}

The morphology of the dust spiral structure can be a key indicator towards the origin of the spirals. We measure symmetric spirals, present in all three wavelengths, with similar radial extension and pitch angles. Symmetric spirals with constant pitch angles are predicted in the case of GI \citep{2018ApJ...860L...5F}, rather than for companion perturbations where asymmetric, variable-pitch angle spirals are expected \citep{2018ApJ...859..119B}. Additionally, we measure similar contrasts for both spirals ($\sim30\%$ difference in contrast between NW and SE spirals), which is also consistent with GI predictions, as companion induced spirals are expected to show a clear primary spiral \citep{2018ApJ...859..119B}. The latter was already shown for the emission at 1.3\,mm \citep{laura_elias, DSHARP_Huang_Spirals}. In this study we extend the finding to 0.89\,mm and 3.3\,mm. Additionally, the spectral index map shows spiral morphology with slightly lower alpha values along the spirals. This coincides with the prediction for dust trapping, expected for GI \citep[][]{2015MNRAS.451..974D}.
 
We measure the optical depth profiles, which appear to show similar values in all three wavelengths. However, the temperature profile used for deriving the optical depth \citep[computed using the flaring and stellar luminosity values from][]{DSHARP_Andrews}, results in a midplane temperature $\sim$2 times higher than the 0.89\,mm brightness temperature at all radii. Most likely, the disk is much colder and optically thick than what we derive, and our optical depths must then be taken as lower limits. 

Several works have shown that when scattering from dust grains is considered, optically thick disks can display lower intensities and be categorized as more optically thin disks \citep{2019ApJ...877L..18Z, 2019ApJ...877L..22L, 2020ApJ...892..136S}. When scattering is not present and the emission is optically thin, then $\alpha$ cannot reach values below 2.0. When dust scattering is included, a region of high optical depth can have $\alpha<2.5$ (and even attain an spectral index below 2.0) if the albedo decreases with wavelength \citep{2019ApJ...877L..18Z, 2020ApJ...892..136S}, something observed in the innermost regions of TW Hya \citep{2016ApJ...829L..35T, 2018ApJ...852..122H}. Furthermore, while $\alpha$ also depends on other dust properties, such as the grain size distribution \citep[e.g.,][]{2014prpl.conf..339T}, values of $\alpha< 3$ are not expected when the optical depth is low, and even in presence of cm-sized grains $\alpha$ should not attain values below $\sim2.5$ \citep{2019ApJ...877L..18Z}. 

From our spectral index profiles $\alpha<2.0$ in the inner $\sim$40\,au and $\alpha<2.5$ inside $\sim$200\,au, outside this region $\alpha$ grows, reaching a maximum value of 3.0 in the outer disk. This indicates that the outer disk is probably optically thin with grains of 0.1-10\,cm, favouring a dust distribution $n(s) \propto s^{-3.5}$ \citep[see Figure 9 in][]{2019ApJ...877L..18Z}. However, in the region between $\sim$70-200\,au the spectral index increases slowly between $\sim$2.2-2.5. This scenario favours a dust distribution $n(s) \propto s^{-2.5}$ and can be explained by either optically thin emission and 0.1-10\,cm or optically thick emission with maximum grain sizes $\sim$0.1 cm. For the inner disk ($\lesssim70$\,au), the spectral index reaches values below 2.0. Most likely, the emission is  optically thick and dust scattering is at work, even when the maximum optical depth we constrain (under standard assumptions) is $\tau \sim 0.5$ at all wavelengths. 
For the DSHARP sample, it has been shown that the optical depths of $\tau \sim 0.5$ at 1.3\,mm in bright rings, can be obtained from optically thick regions with a scattering albedo of $\omega_{\nu} \sim$0.89 \citep{2019ApJ...877L..18Z}. 
For Elias 2-27, we measure an average $\tau \sim 0.45$ at 1.3\,mm inside 70\,au, which could be obtained with a scattering albedo of $\omega_{\nu} \sim$ 0.93 \citep[see equations 14 and 15 from ][]{2019ApJ...877L..18Z}. This albedo value is sufficient to mask optically thick 1.3\,mm emission ($\tau_{real} \sim$1-5) in the inner disk, which we would be inferring as optically thin 1.3\,mm emission ($\tau_{obs} \sim 0.45$) in the standard assumption of only absorption opacity. As was previously discussed, our optical depths values could be underestimated due to the difference between the measured brightness temperatures and model midplane temperature. If the midplane temperature was a factor of $\sim$1.5 lower, the optical depth in the inner 70\,au of the disk, at 1.3\,mm would be $\tau \sim 0.99$ which could be obtained with a scattering albedo $\omega_{\nu} \sim$ 0.74. Similar analysis can be done with the other wavelengths from the measured $\tau$ values of the inner disk. This results in a scattering albedo of 0.92 and 0.95 for 0.89\,mm and 3.3\,mm respectively.

Together with the effect of scattering at long wavelengths, the observed low spectral index values can also occur due to the disk's temperature. For cold ($<$30K) systems, such as Elias 2-27, spectral indices below 2 are expected due to the displacement of the peak blackbody radiation to the sub-mm range \citep{2020ApJ...892..136S}. Additionally, if the emission is optically thick and there are fluctuations in the vertical temperature structure, this could also result in $\alpha<2.0$ \citep[see Figure 5 in][]{2020ApJ...892..136S}. 

The underestimation of the optical depth in Elias 2-27 when not including the effect of scattering will impact its solid mass estimate. This effect is larger for inclined disks and when the emission area is compact \citep{2019ApJ...877L..18Z}, which is the case of our source in the inner regions. Considering the large disk extent (up to $\sim250$\,au), most of the dust mass resides in the optically thin outer disk, thus, the disk mass could be underestimated by up to a factor of $\sim$2 \citep{2019ApJ...877L..18Z}. The latter implicates that the previously contrained disk-to-star mass ratio of 0.1-0.3\citep[][using standard assumptions such as a gas-to-dust ratio of 100]{2009ApJ...700.1502A, 2010A&A...521A..66R} is a lower bound. If the disk mass was higher by up to a factor of 2, the resulting disk-to-star mass ratio would make gravitational instabilities a likely cause for the spiral structure.

\subsection{Asymmetries and Perturbations in the Gas}

The highly perturbed morphology constrained for the emitting gas layer in Elias 2-27 is a new characteristic for this system and offers new insight into the ongoing dynamic processes. The asymmetric structure of the $^{13}$CO and C$^{18}$O emitting layer, as well as the dust spiral arms present in Elias 2-27, could be in principle caused by fly-by interaction \citep[e.g., ][]{2019MNRAS.483.4114C} or a external companion. But if this were the case, we also expect a strong kinematical perturbation in the integrated emission and velocity maps of Elias 2-27, such as those reported by \cite{DSHARP_Nico} for disks with known external companions. Furthermore, observations in near-infrared of Elias 2-27 have not found any companion \citep{2009ApJ...696L..84C, 2020MNRAS.tmp.2019Z}.

With no outer perturbation, the emission layer height of the gas should follow hydrostatic equilibrium and its structure depend on the gas temperature \citep{2015arXiv150906382A}. In this case we expect a cone-like or flared emission layer, with height increasing at larger radial distance \citep{2013ApJ...774...16R}. We will discuss two possible origins for the observed asymmetries in the gas: ongoing infall of material from its surrounding cloud/envelope, or a warped inner disk causing azimuthal temperature variations.    

Three-dimensional simulations of circumstellar disks with ongoing accretion show that the vertical structure of the disk will become asymmetric, as the accreting gas shocks the disk from above or below, along the $z$ plane, as described by cylindrical coordinates \citep[see Figure 8 in][]{2017A&A...599A..86H}. Furthermore, simulations for ongoing infall predict the appearance of spiral structures in the surface, generated by the infall process and shocks, both in 3D simulations \citep{2017A&A...599A..86H, 2011MNRAS.413..423H} and 2D simulations \citep{2015A&A...582L...9L}. Infall-triggered spirals may have been observed in the 343\,GHz ALMA emission of VLA 1, a Class I source with active envelope infall \citep[][]{2020NatAs...4..142L}. From our observations, we have shown the presence of large-scale emission surrounding the disk (see Figures \ref{mom_largescale}, \ref{panel_largescale_13CO} and \ref{panel_largescale_C18O}). Our data lacks appropriate uv-coverage to accurately sample the whole FOV, which extends for 20$\arcsec$, this is possibly the cause of the striped pattern seen in the channel maps (Figures \ref{panel_largescale_13CO} and \ref{panel_largescale_C18O}). We do not recover velocity gradients connecting the large-scale emission to the disk emission, this could be due to lack of sensitivity or angular resolution, nevertheless, it opens the option for infall to be ongoing. While this is not expected for a Class II source, it has been proposed that Elias 2-27 could be a very young Class II disk \citep{2017ApJ...835L..11T}. 

An azimuthal variation of the temperature in the disk could also explain an azimuthally varying emission layer height, which can be expected when a disk is warped. A warp will affect the disk illumination, depending on the position and characteristic angles of the warp itself \citep{2010MNRAS.403.1887N}. Warps are generally detected through their shadowing effects in scattered light \citep[]{2015ApJ...798L..44M, 2018A&A...619A.171B} and their distinct kinematical signatures \citep[]{2017MNRAS.466.4053J, 2020ApJ...889L..24P, 2017A&A...607A.114W}. Given its high extinction, no scattered light observations are available for Elias 2-27, so we can only compare our observations with the kinematic predictions for a warp. 

In ALMA observations, a misaligned disk can be inferred through the kinematical signature in the velocity map: if a non-misaligned disk is modeled then positive and negative residuals appear, opposite with respect to a ``symmetry'' axis \citep[e.g., as seen in the velocity field residuals for HD100546 or HD143006, ][]{2017A&A...607A.114W,DSHARP_Laura_10}. We do not observe such residuals in our data (see Figure \ref{kep_model_vel}), rather we see an ``X'' shaped residual in the C$^{18}$O emission. Other signatures related to warps are asymmetric illumination, deviations in line profiles, and twisted features in the channel and integrated emission maps \citep{2017MNRAS.466.4053J, 2018MNRAS.473.4459F}. The gas emission of Elias 2-27 is characterized by a strong illumination asymmetry between North-South sides of the disk. Additionally, we have shown the presence of ``curved'' and ``wavy'' features and deviations in both the integrated velocity maps and velocity cube (see Figures \ref{kep_model_vel} and \ref{kinks_central} respectively). These features may relate to an inclined disk, warped such that the disk bends perpendicular to the line of sight \citep[see Case C in ][]{2017MNRAS.466.4053J}. The latter configuration produces channel maps with curved structures and asymmetric illumination, such as observed in our data, while it also shows further structure in the integrated intensity map and asymmetric line profiles \citep{2017MNRAS.466.4053J}. We may not be sensitive to all these features given our moderate spatial resolution (simulations are done with resolution $\sim$ 0.1$\arcsec$, we have $\sim$ 0.3$\arcsec$) and the effects of cloud contamination in the system. 

If indeed there is a warp, we should be able to roughly trace its location through the temperature variations it will cause. We expect temperature variations to affect the emitting layer's height and should therefore be able to trace these variations in the projected emission. While we do observe an elevation difference in the East and West sides of the disk, apparently separated by the semi-minor axis, we have discussed that this is probably by chance and that we cannot determine an exact symmetry angle. The method used to derive the emitting surface assumes symmetry with respect to the semi-major axis and this biases our results. We also show that, when tracing the deprojected border of emission (see Figure \ref{C18O_gasgap}) there are two local maxima and minima radial extensions, suggesting that there is not in fact a single symmetry axis \citep[we would have expected only one minimum and maximum extent, roughly symmetrically opposed for the case of a warp,][]{2017MNRAS.466.4053J}. More complex processes, or a combination of effects are occurring. 

Finally, if a warp is responsible for the kinematic effects, there is the question of its origin. On one hand, warps are thought to arise from close-in binary interactions or inclined planetary orbits \citep[]{2018MNRAS.481...20N, 2020MNRAS.492.3306A}. Warps in very young systems, possibly produced by the infall of material, have been predicted \citep{2010MNRAS.401.1505B} and also reported \citep{2019Natur.565..206S}. In the case of Elias 2-27 we have discussed that infall may be ocurring, given the detection of large-scale emission at velocities close to the disk velocities. If the disk is indeed warped, this effect could be caused either by a planet in an inclined orbit or infall, both options require further investigation.

Aided by the isovelocity curves computed from the constrained emission surface of the disk, we detect multiple deviations from Keplerian motion in the velocity cube of $^{13}$CO and C$^{18}$O, co-located with the spiral structure. The isovelocity curves themselves show a perturbed nature, given the complex emission surface. The colocation of the perturbations with the spiral structures and the strong deviations could indicate a connection between the spirals and the emitting surface morphology. Previously reported deviations for planetary companions have been of around 15$\%$ \citep{2018ApJ...860L..13P} with respect to the channel velocity, while the perturbations we observe are much larger, reaching up to 80$\%$ of the channel velocity for some of the features in the central channel maps. Such large perturbations increase the likelihood that they relate to the spirals, rather than to a companion. Large planetary deviations require a high perturber mass, and in that case we would expect to see its effect in the dust emission. Furthermore,  we expect deviations caused by a planet to be spatially localized \citep[]{2018ApJ...860L..13P, 2020ApJ...890L...9P, 2019NatAs...3.1109P} and our observations show deviations on both East and West sides of the disk, present along several channels. The detected perturbations agree with the predictions of \cite{2020arXiv200715686H}, they are co-located with the spirals and the morphology of the kink is similar to what was predicted in their work. If indeed the disk is warped or suffering considerable infall of material, as previously discussed, the observed ``kinks'' could be the combination of both kinematical deviations induced by a warp and perturbations due to the spirals.

In this study we attempted a purely Keplerian fit to the expected super-Keplerian velocity profile when GI is the governing process \citep{1999A&A...350..694B, 2007NCimR..30..293L}. The Keplerian rotation curve is able to fit the observed velocity profile, but we do note that the East channels, especially from the C$^{18}$O emission, show super-Keplerian velocities. Further analysis is needed to check if a self-gravitating rotation curve may be a better description of this data. This is currently being studied and will be presented in a future publication (Veronesi et al, submitted).

Finally, regarding the gap in the C$^{18}$O gas emission, we do not see evidence of it being produced by a physical perturber. The gap does not appear to be colocated with any perturbation in the channel maps (see Figure \ref{kinks_central}), which we would expect if the gap origin was planetary \citep[e.g.][]{2018ApJ...860L..13P, 2018ApJ...860L..12T}. It may answer to chemical processes of the gas or optical depth effects \citep[see discussion in ][]{2018ApJ...869L..48G}.

\subsection{Comparison with SPH simulations.}

We find that gravitational instabilities can accurately reproduce the spiral morphology at multiple wavelengths, with parameters close to the observational constraints, as shown in Figure \ref{model_q03_98}. However we see considerable morphological differences in the comparison to high-angular resolution data. The width and morphology of spiral arms in GI environments can be regulated by varying the cooling parameters \citep[see figures from ][]{2003MNRAS.339.1025R, 2017ApJ...836...53B}. Future SPH simulations should attempt to sample this parameter space to further understand the cooling prescription of the disk. We note that \citeauthor{2020MNRAS.498.4256C} also analyses the spiral structure of Elias 2-27 at DSHARP angular resolution and they obtain thicker spirals in the substracted image. These spirals are constructed with a semi-analytical model, considering grain growth and while they are thicker than what we recover, they are still not able to reproduce the precise morphology of the  observations, as they are too smooth \citep[see substracted images in Figure 16 from ][]{2020MNRAS.498.4256C}. 

GI does not explain the dust gap, if the dust gap was carved by a planet, estimates indicate at best a mass of 0.1M$_J$ \citep{DSHARP_Zhang}, which is much smaller than planets expected to be formed in a GI environment. Though not shown in this work, we produced additional SPH simulations of an unstable disk with different planetary-mass companions. In all cases, after a few orbits, the planet migrated onto the star. This is expected, according to the predictions of Type I migration for planets orbiting at several au from the star \citep{2014prpl.conf..667B}. Not allowing migration could allow a gap to form \citep[see ][]{2017ApJ...839L..24M}, however it would not be a realistic scenario. 

In our spectral line observations we recover perturbations that are colocated with the spirals and span several channels, sharing some similarities to those predicted by for GI disks \citet{2020arXiv200715686H}. Our gas simulations from the best-fit GI parameters do not show the perturbation features from \citet{2020arXiv200715686H} when sampled and imaged with the uv-coverage of the observations. This implies that the inner perturbation observed in the $^{13}$CO and C$^{18}$O channel maps is indeed very large, as it appears at our lower angular resolution. The perturbation is seen across several channels, which is in agreement with predicted perturbations of a GI disk \citep{2020arXiv200715686H}. The fact that we don't see the perturbation in the channel maps from simulated emission could be due to the decisions regarding the chosen time-stamp for the simulation and the position angle. As we select a specific time-frame, it is possible that at other evolutionary stages, the strength of the gas perturbations would have been larger. Also, it has been shown by \cite{2020arXiv200715686H} that the strength of the  ``GI-wiggle'' will vary depending on the position angle of the emission. While the inclination of the simulation matches the value of the observations (56.2$^\circ$), the position angle is set before inclining the disk. When producing the ALMA mock image, the position angle of the emission is determined by the value that allows a good visual comparison of the final, inclined, simulated emission image to the observations. Due to the computational expense of sampling various position angles, we adopted the observational value (118.8$^\circ$)  for the position angle of the simulations, varying only by 90$^\circ$ or 180$^\circ$ in some cases. These shifts were decided using a visual criteria. We therefore note that there could be a range of position angle values that allow a good comparison with the observations, while also producing stronger kinematic perturbations, this is not tested in this work.

\subsection{Spiral Structure Origin}

The hypothesis for the origin of the spiral arms observed in the dust emission of Elias 2-27 are either gravitational instabilities or perturbation by a companion. From the observational constraints presented in this work there are some key features that may help us define which scenario fits best the Elias 2-27 disk. To begin, however, we must state that neither option can accurately predict, on its own, both the spirals and the dust gap. In the case of a disk undergoing GI, gaps are not expected features even if a planet was formed, given the fast migration even massive planets will have under these conditions \citep{2011MNRAS.416.1971B}. In order to reproduce the spiral arms a companion would have to be located beyond the spiral extent \citep{2017ApJ...839L..24M}. The possibility of the spirals being formed by a companion seems unlikely, as has been discussed in previous studies, given the contrast, symmetry, and extent of the spirals \citep{2018ApJ...860L...5F, 2018ApJ...859..119B}. Furthermore, the possibility of an external perturber, such as a stellar companion or a fly-by causing the spiral structure, is also unlikely, due to the lack of a clear kinematical signature in the data and the non-detection of any object nearby \citep{2005A&A...437..611R, 2009ApJ...696L..84C, 2020A&A...635A.162L}. \cite{2020A&A...635A.162L} conducted a NACO/VLT survey in search for planetary companions around 200 stars, including Elias 2-27. For this system they reach a 50$\%$ probability of detecting a 2$M_J$ companion outside 100\,au or a $\sim$10$M_J$ companion at 40-50\,au \citep[][R. Launhardt, priv. comm.]{2020A&A...635A.162L}. We note that the detection of an external perturber is made more challenging by the extinction that affects this region.

Elias 2-27 has been shown to have a large disk-to-star mass ratio \citep{2009ApJ...700.1502A, 2010A&A...521A..66R}, in this work we additionally discuss the possibility of the disk mass being up to a factor of $\sim$2 higher, if scattering is a relevant process \citep{2019ApJ...877L..18Z}. Even if scattering wasn't relevant, we show that the disk is probably more optically thick than the reported values, which also points towards a larger disk mass. The values of disk-to-star mass ratio is sufficient to excite gravitational instabilities and we can accurately reproduce the spiral morphology using SPH models of a self-gravitating disk in the medium resolution at least. Additionally, we detect dust trapping signatures in the continuum observations, in the contrast variations with increasing wavelength and lower spectral index values along the spirals.  We also measure strong kinematic perturbation co-located with the spirals over multiple channels. The high disk mass, together with the strong deviations from Keplerian motion, consistent with the kinematical prediction for a GI disk \citep{2020arXiv200715686H}, lead us to signal the origin of the spirals to be gravitational instabilities, rather than a companion.

Infall of material would explain the high disk-to-star mass ratio of the system and the excitation of spiral structures due to GI \citep{2017ApJ...838..151T, 2017A&A...599A..86H}. While infall mechanisms are expected to be present in younger Class 0/I systems than Elias 2-27, which is classified as a Class II disk \citep{2009ApJ...700.1502A}, it has been proposed that Elias 2-27 could be an extremely young Class II object to explain the spiral structure \citep{2017ApJ...835L..11T}. Though GI can explain the spiral structure traced in the dust, there is also a clear dust gap \citep{DSHARP_Huang_Annular}, which emphasize is not a feature predicted or explained by GI. On the other hand, infall may explain on its own the perturbed morphology of the gas emission layer, but there is also a marked brightness asymmetry, which could be related to the presence of a warp in the disk \citep{2010MNRAS.403.1887N}. If the disk is warped, due to infalling material breaking the disk, this could possibly explain the dust gap observed at high angular resolution by \cite{DSHARP_Huang_Annular}, if the separation between inner and outer misaligned disks were located at $\sim$70\,au from the central star. Certainly further observations on the source are required, to sample shorter baselines and adequately study the dynamics of the large-scale emission, searching for infall signatures, and also observations at higher spatial resolution, to analyze the origin of the dust gap and better constrain the kinematic perturbations in the system. 

\newpage
\section{Summary}

We have presented and analyzed multi-wavelength dust continuum observations of the protoplanetary disk around Elias 2-27. We also studied the gas emission for $^{13}$CO and C$^{18}$O $J=3-2$. This provides new observational constraints on this source, which allowed us to study the origin of the prominent spiral structure. Our findings are as follows:

\begin{itemize}
    \item The spiral substructure is present in dust observations at multiple wavelengths, from 0.89 to 3.3\,mm, and shows a higher contrast at longer wavelengths. These signs are possible indicators of grain growth and dust trapping at the spiral arm location. 
    
    \item From the spectral index analysis we trace a spiral morphology with lower spectral index values along the spiral location. This is expected for dust trapping, which is a key signature of gravitational instabilities not observed before in other systems with spiral morphology. The spectral index values also indicate the presence of large grains in the outer disk. Inwards of $\sim 70$au the spectral index drops to values even lower than 2. This can be explained if this region has high optical depth and high albedo in the presence of dust scattering. We discuss that our optical depth estimates are lower limits, given the low brightness temperature measured in the observations. The presence of scattering and higher optical depths implies that the solids mass estimated under standard assumptions is likely a lower limit. The mass of the disk in Elias 2-27 is presumably higher than previously estimated.
    
    \item We compute SPH simulations of a gravitationally unstable disk with parameters as those in Elias 2-27, and are able to replicate the spiral morphology at the three different wavelengths we study at $\sim$0.2$\arcsec$ resolution. Discrepancies at high angular resolution could be due to the cooling prescription used.
    
    \item Observations show that the gas emission is not azimuthally symmetric in the vertical direction, i.e. the disk has a larger emission layer height in the West than in the East at most radial distances. Additionally, at larger radial distances, the kinematic data indicate that the emission layer height decreases. This is the first time we observe an azimuthal emission layer height difference in a protoplanetary disk, and it does not appear in predictions for either a GI disk or the presence of a planetary companion.
    
    \item Tracing the different heights of the $^{13}$CO and C$^{18}$O emission layers we show that $^{13}$CO comes from a higher layer than C$^{18}$O, with velocities consistent with Keplerian rotation. The stellar mass we constrain ($\sim$0.46-0.5 M$_\odot$), is in agreement with the literature value (0.49M$_\odot$). Gas emission depletion (a gap) is observed in the distribution of  C$^{18}$O at a radius of $\sim$240\,au. This gap does not appear to be co-located with the main perturbations we recognize in the channel maps.

    \item We see ``kinks'' or perturbations in the channel maps of both CO tracers, that appear co-located with the spiral features. These kinks are stronger and present across a wide velocity range, making it unlikely that they have a planetary origin. The characteristics of these perturbations are similar to what has been predicted in a GI disk by \cite{2020arXiv200715686H}. 
    
    \item Based on observations that show large-scale emission surrounding and connecting to the disk, we propose the infall of material from the surrounding cloud is responsible for exciting GI in the disk, which in turn causes the dust spiral arm features. Infall of material can also explain the perturbed emission layer constrained from the gas tracers. Additionally, if infall warped the disk, this could explain the brightness asymmetry in the channel and integrated emission maps. Depending on the warp location it could also explain the dust gap observed at higher angular resolution. Further observations are necessary to effectively detect the presence of a warp and confirm ongoing infall. \\

\end{itemize}

\software{CASA \citep{2007ASPC..376..127M}, eddy \citep{2019JOSS....4.1220T}, bettermoments \citep{2019ascl.soft01009T}, PHANTOM \citep{2018PASA...35...31P}, mcfost \citep{mcfost1, mcfost2}, galario \citep{2018MNRAS.476.4527T}, frankenstein \citep{2020MNRAS.495.3209J}, Astropy \citep{astropy:2013, astropy:2018},
Matplotlib \citep{Hunter:2007}, emcee \citep{2013PASP..125..306F}}

\begin{acknowledgements}
This paper makes use of the following ALMA data: \#2013.1.00498.S, \#2016.1.00606.S and  \#2017.1.00069.S. 
ALMA is a partnership of ESO (representing its member states), NSF (USA), and NINS (Japan), together with NRC (Canada),  NSC and ASIAA (Taiwan), and KASI (Republic of Korea), in cooperation with the Republic of Chile. 
The Joint ALMA Observatory is operated by ESO, AUI/NRAO, and NAOJ. 
L.M.P.\ acknowledges support from ANID project Basal AFB-170002 and from ANID FONDECYT Iniciaci\'on project \#11181068.
M.B.\ acknowledges funding from ANR of France under contract number ANR-16-CE31-0013 (Planet Forming Disks). 
C.H.\ was a Winton Fellow and this research was supported by Winton Philanthropies / The David and Claudia Harding Foundation. A.S.\ acknowledges support from ANID/CONICYT Programa de Astronom\'ia Fondo ALMA-CONICYT2018 31180052.
J.M.C.\ acknowledges support from the National Aeronautics and Space Administration under grant No. 15XRP15\_20140 issued through the Exoplanets Research Program. 
S.M.A.\ acknowledges funding support from the National Aeronautics and Space Administration under Grant No. 17-XRP17 2-0012 issued through the Exoplanets Research Program.
J.B.\ acknowledges support by NASA through the NASA Hubble Fellowship grant \#HST-HF2-51427.001-A awarded  by  the  Space  Telescope  Science  Institute,  which  is  operated  by  the  Association  of  Universities  for  Research  in  Astronomy, Incorporated, under NASA contract NAS5-26555.
Th.H.\ acknowledges support from the European Research Council under the 
Horizon 2020 Framework Program via the ERC Advanced Grant Origins 83 24 28.
L.L.\ acknowledges the financial support of DGAPA, UNAM (project IN112820), and CONACyT, México. 
M.T.\ has been supported by the UK Science and Technology research Council (STFC) via the consolidated grant ST/S000623/1.
L.T.\ acknowledges support from the Italian Ministero dell Istruzione, Universit\`a e Ricerca through the grant Progetti Premiali 2012 – iALMA (CUP C$52$I$13000140001$), by the Deutsche Forschungs-gemeinschaft (DFG, German Research Foundation) - Ref no. FOR $2634$/$1$ TE $1024$/$1$-$1$, and the DFG cluster of excellence Origins (www.origins-cluster.de).
This project has received funding from the European Union's Horizon 2020 research and innovation programme under the Marie Sklodowska-Curie grant agreement No 823823 (DUSTBUSTERS) and from the European Research Council (ERC) via the ERC Synergy Grant {\em ECOGAL} (grant 855130). 
Powered@NLHPC: This research was partially supported by the supercomputing infrastructure of the NLHPC (ECM-02). 
This research used the ALICE2 High Performance Computing Facility at the University of Leicester. This research also used the DiRAC Data Intensive service at Leicester, operated by the University of Leicester IT Services, which forms part of the STFC DiRAC HPC Facility (\url{www.dirac.ac.uk}). The equipment was funded by BEIS capital funding via STFC capital grants ST/K000373/1 and ST/R002363/1 and STFC DiRAC Operations grant ST/R001014/1. DiRAC is part of the National e-Infrastructure. This work was partially supported by the University of Georgia Office of Research and the Department of Physics and Astronomy.

\end{acknowledgements}

%






\appendix

\section{Additional Gas Analysis Figures}

\begin{figure*}
   \centering
   \includegraphics[width=\hsize]{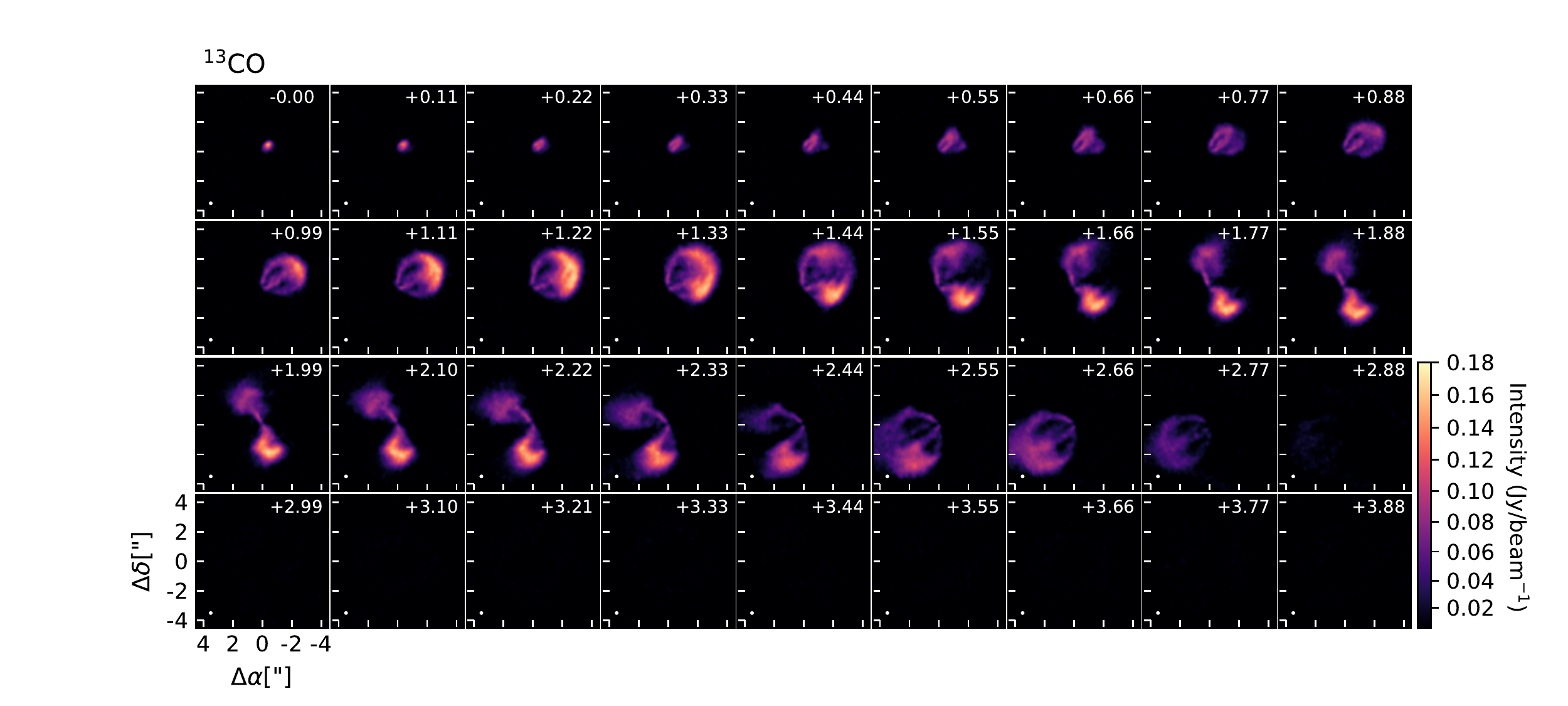}
      \caption{Individual emission of each channel for $^{13}$CO emission, corresponding velocities are written in the top right of each panel.
              }
         \label{panel_13CO}
\end{figure*} 

\begin{figure*}
   \centering
   \includegraphics[width=\hsize]{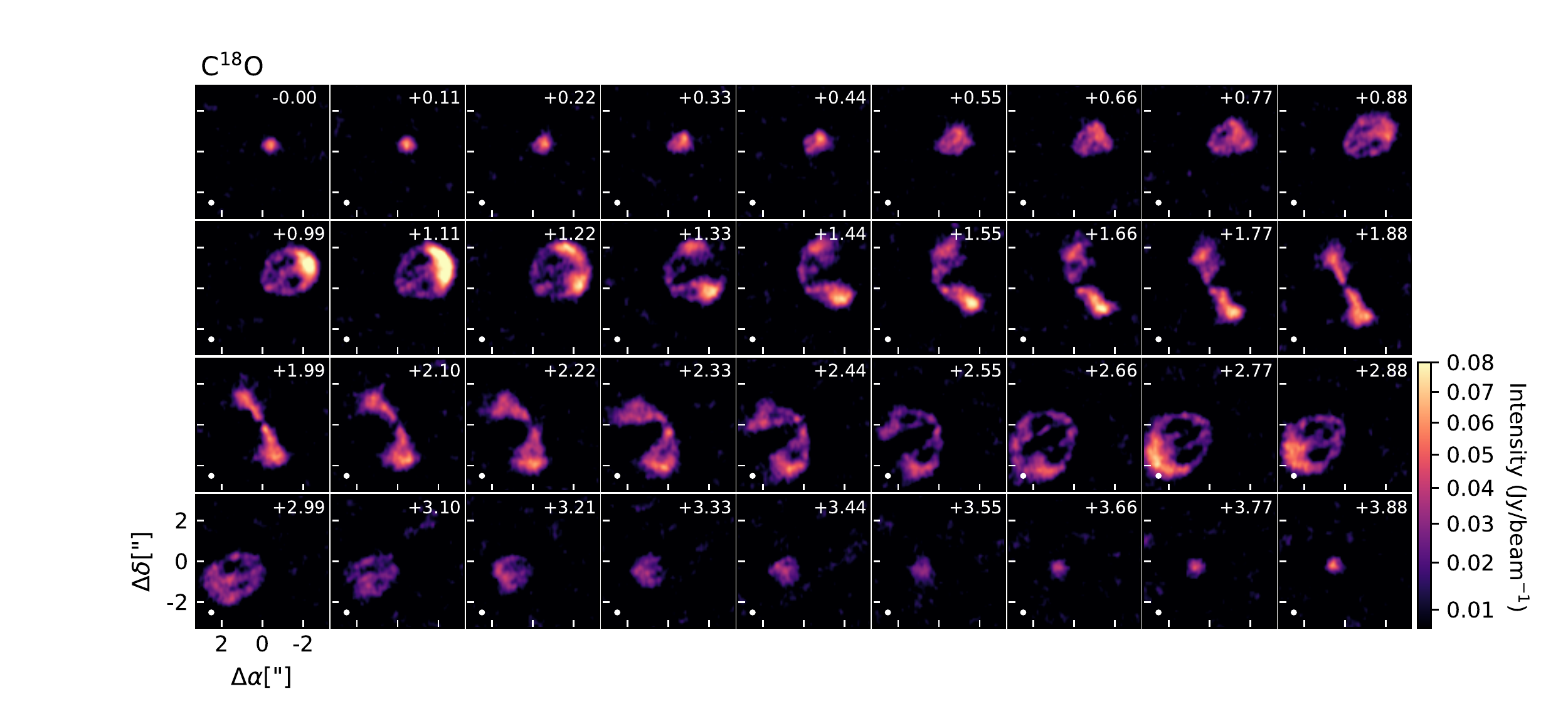}
      \caption{Individual emission of each channel for C$^{18}$O emission, corresponding velocities are written in the top right of each panel.
              }
         \label{panel_C18O}
\end{figure*} 

\begin{figure*}
   \centering
   \includegraphics[width=\hsize]{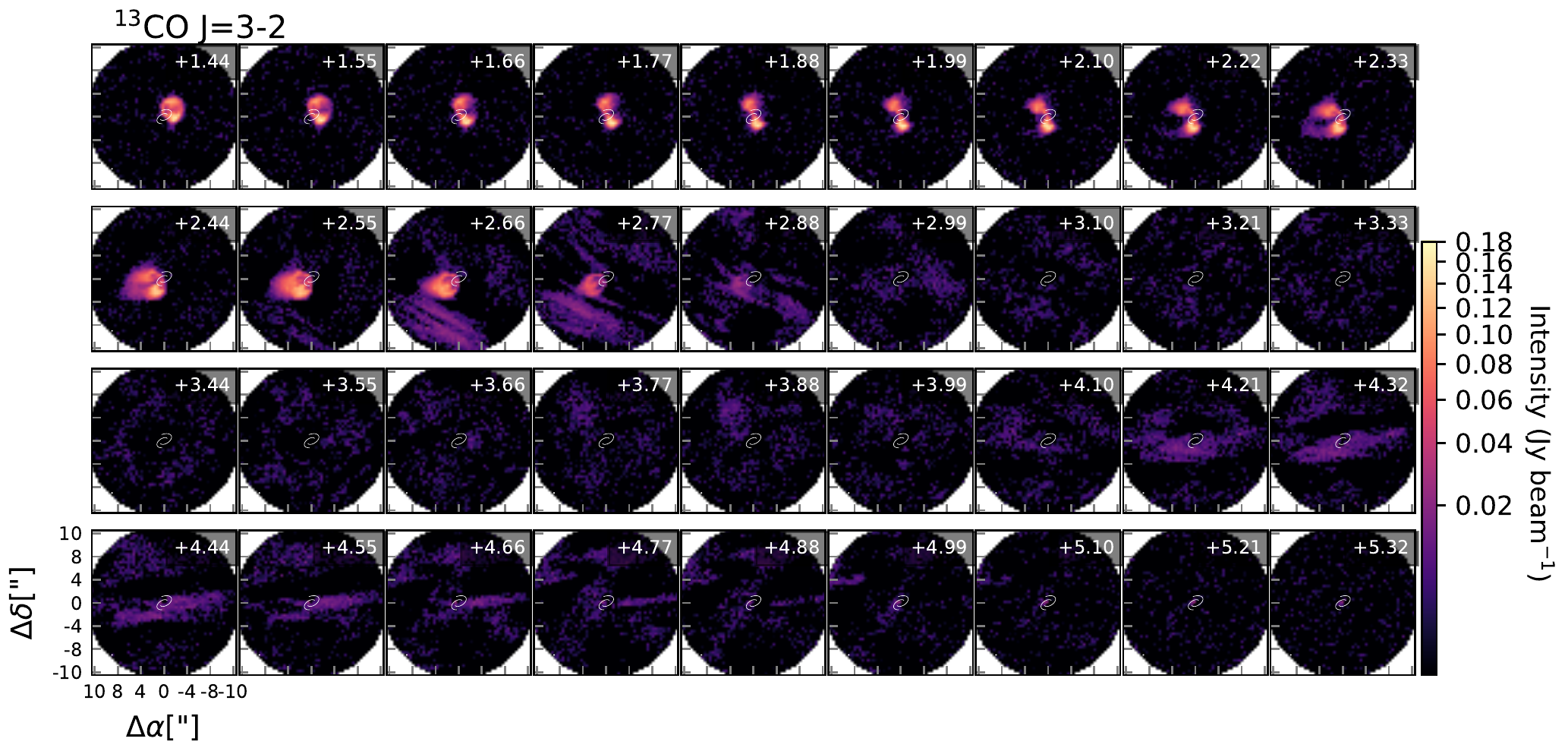}
      \caption{Individual emission of each channel for $^{13}$CO emission, corresponding velocities are written in the top right of each panel. Channels correspond to velocities were large-scale emission is observed. Imaging enhances large-scale structure, parametric model of the spiral arms is plotted for reference.
              }
         \label{panel_largescale_13CO}
\end{figure*} 

\begin{figure*}
   \centering
   \includegraphics[width=\hsize]{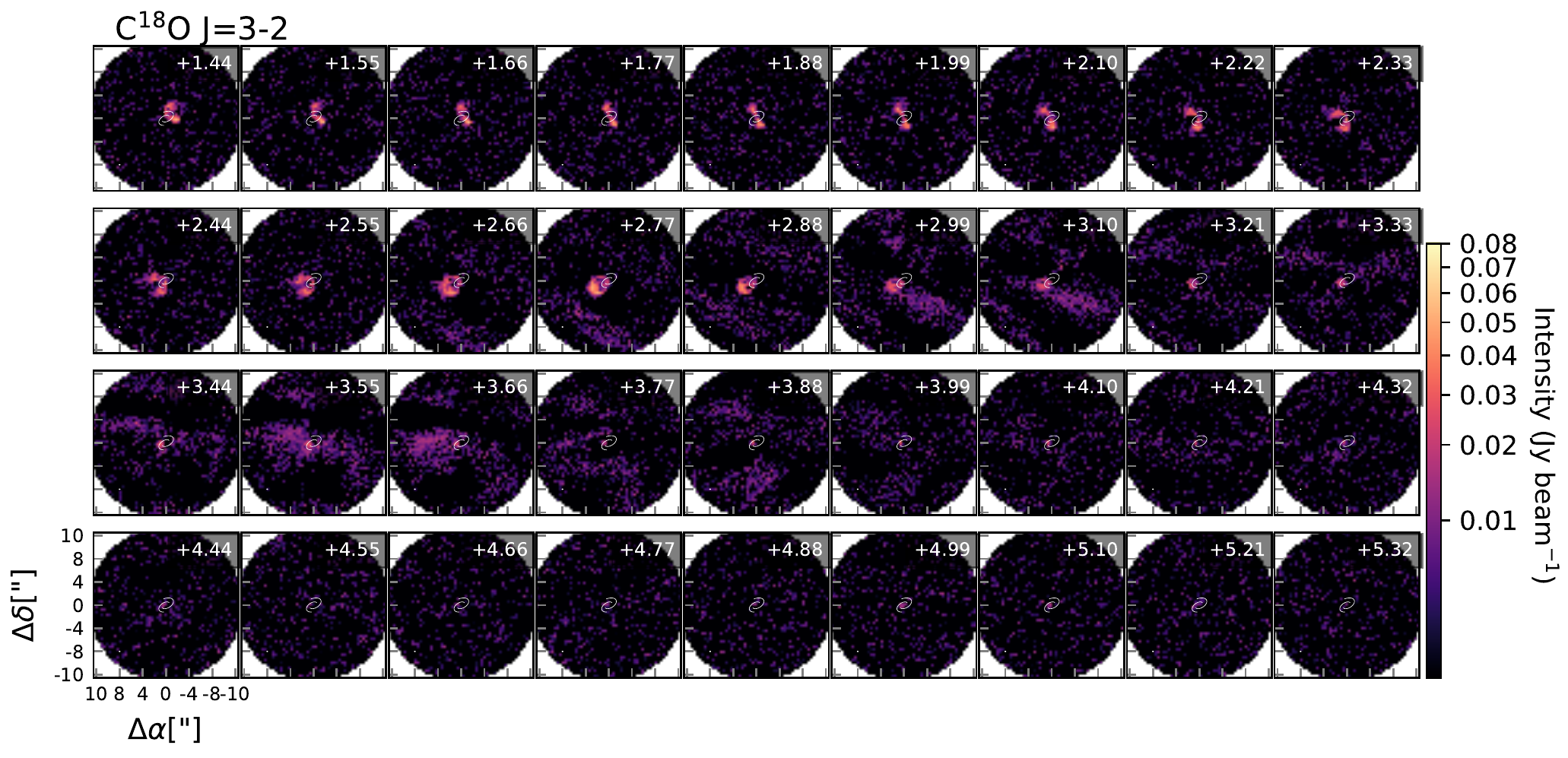}
      \caption{Individual emission of each channel for C$^{18}$O emission, corresponding velocities are written in the top right of each panel. Channels correspond to velocities were large-scale emission is observed. Imaging enhances large-scale structure, parametric model of the spiral arms is plotted for reference.
              }
         \label{panel_largescale_C18O}
\end{figure*}

\begin{figure}
   \centering
   \includegraphics[width=\hsize]{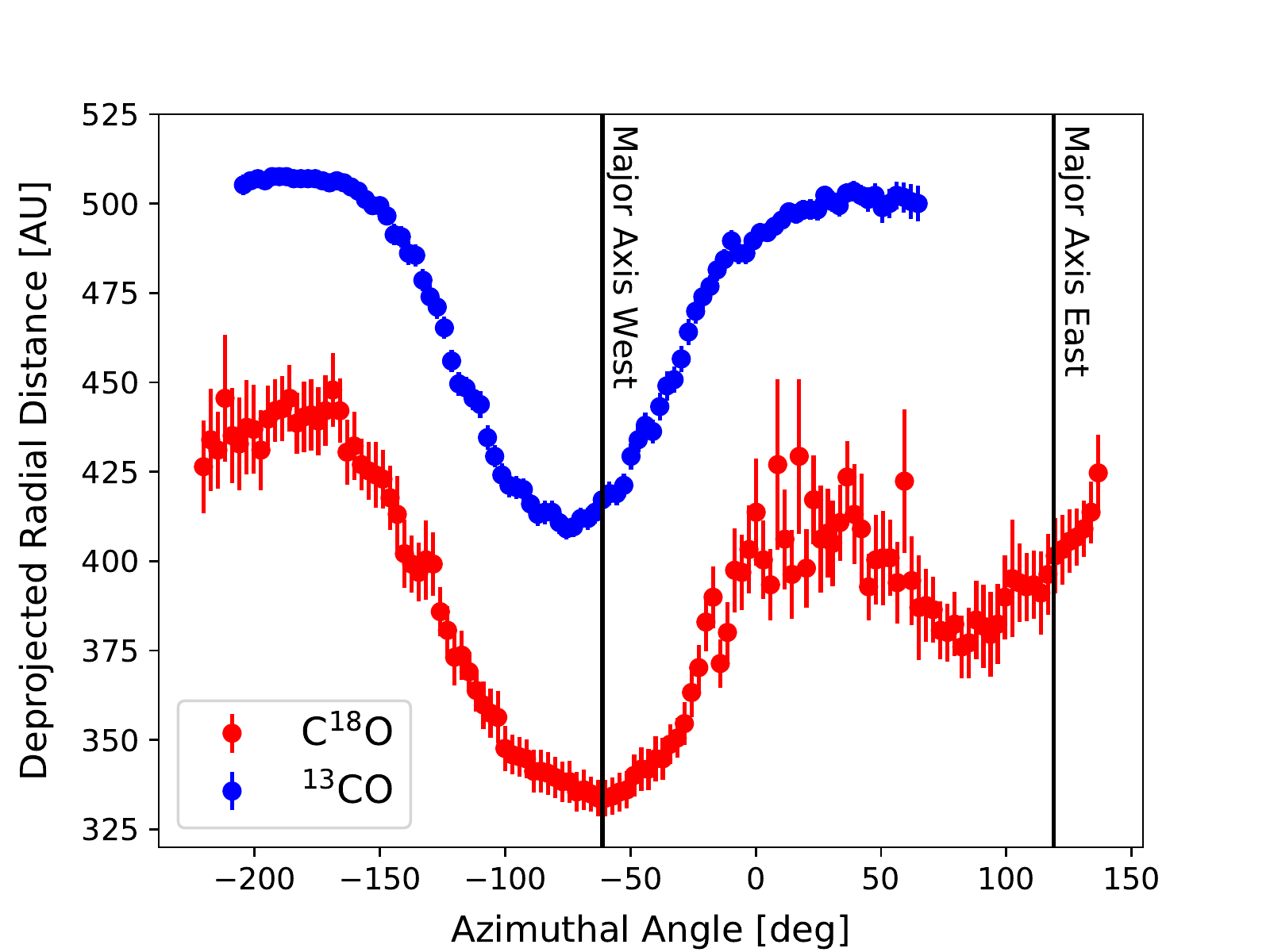}
      \caption{Radial extension of the integrated emission map for $^{13}$CO (in blue) and C$^{18}$O (in red), obtained for each azimuthal angle as the radius that encloses 98$\%$ of the as the azimuthally integrated emission.
              }
         \label{border_13CO_C18O}
\end{figure}

\section{Additional SPH Model Details}

\begin{rotatetable*}
\begin{deluxetable*}{c c c c c c c c}
\tablecaption{Pitch Angle values for Simulations}

\tablehead{
\colhead{Simulation} & \colhead{NW$_{0.89\,mm}$} & 
\colhead{SE$_{0.89\,mm}$} & \colhead{NW$_{1.3\,mm}$} & 
\colhead{SE$_{1.3\,mm}$} & \colhead{NW$_{3.3\,mm}$} & 
\colhead{SE$_{3.3\,mm}$} & \colhead{likelihood $\phi$}
} 
\startdata
tapered: 0.7   &  &  &  &  &  &  & \\ 
q = 0.2        &  13.46$^{\circ}$ $\pm$ 0.08$^{\circ}$ & 14.96$^{\circ}$ $\pm$ 0.08$^{\circ}$ & 10.86$^{\circ}$ $\pm$ 0.06$^{\circ}$ & 15.27$^{\circ}$ $\pm$ 0.08$^{\circ}$ & 9.85$^{\circ}$ $\pm$ 0.07$^{\circ}$ & 13.7$^{\circ}$ $\pm$ 0.1$^{\circ}$ & 29.31\\ 
q = 0.3        & 12.52$^{\circ}$ $\pm$ 0.06$^{\circ}$ & 11.54$^{\circ}$ $\pm$ 0.06$^{\circ}$ & 12.24$^{\circ}$ $\pm$ 0.06$^{\circ}$ & 10.82$^{\circ}$ $\pm$ 0.05$^{\circ}$ & 12.09$^{\circ}$ $\pm$ 0.06$^{\circ}$ & 10.6$^{\circ}$ $\pm$ 0.06$^{\circ}$  & 24.04\\
q = 0.5        & \nodata & \nodata & \nodata & \nodata & \nodata & \nodata & \nodata \\
tapered: 1.0   &  &  &  &  &  &  & \\ 
q = 0.2        & \nodata & \nodata & \nodata & \nodata & \nodata & \nodata & \nodata  \\ 
q = 0.3        & 12.75$^{\circ}$ $\pm$ 0.06$^{\circ}$ & 11.42$^{\circ}$ $\pm$ 0.05$^{\circ}$ & 12.34$^{\circ}$ $\pm$ 0.06$^{\circ}$ & 11.33$^{\circ}$ $\pm$ 0.05$^{\circ}$ & 13.08$^{\circ}$ $\pm$ 0.06$^{\circ}$ & 11.19$^{\circ}$ $\pm$ 0.05$^{\circ}$ & 21.04 \\
q = 0.5        & \nodata & \nodata & \nodata & \nodata & \nodata & \nodata & \\
power law: 1.3   &  &  &  &  &  &  & \\ 
q = 0.2        & 13.9$^{\circ}$ $\pm$ 0.08$^{\circ}$ & 11.16$^{\circ}$ $\pm$ 0.11$^{\circ}$ & 14.75$^{\circ}$ $\pm$ 0.07$^{\circ}$ & 10.40$^{\circ}$ $\pm$ 0.08$^{\circ}$ & 14.82$^{\circ}$ $\pm$ 0.07$^{\circ}$ & 10.48$^{\circ}$ $\pm$ 0.11$^{\circ}$ & 36.19\\ 
q = 0.3        & 14.88$^{\circ}$ $\pm$ 0.07$^{\circ}$ & 15.66$^{\circ}$ $\pm$ 0.07$^{\circ}$ & 15.53$^{\circ}$ $\pm$ 0.06$^{\circ}$ & 15.36$^{\circ}$ $\pm$ 0.07$^{\circ}$ & 15.34$^{\circ}$ $\pm$ 0.07$^{\circ}$ & 15.92$^{\circ}$ $\pm$ 0.06$^{\circ}$  & 45.32\\
q = 0.5        & 14.86$^{\circ}$ $\pm$ 0.07$^{\circ}$ & 18.08$^{\circ}$ $\pm$ 0.07$^{\circ}$ & 14.47$^{\circ}$ $\pm$ 0.06$^{\circ}$ & 17.74$^{\circ}$ $\pm$ 0.08$^{\circ}$ & 14.2$^{\circ}$ $\pm$ 0.06$^{\circ}$ & 18.82$^{\circ}$ $\pm$ 0.07$^{\circ}$ & 53.08\\
power law: 1.5   &  &  &  &  &  &  & \\ 
q = 0.2        & 10.94$^{\circ}$ $\pm$ 0.09$^{\circ}$ & 7.34$^{\circ}$ $\pm$ 0.06$^{\circ}$ & 12.38$^{\circ}$ $\pm$ 0.09$^{\circ}$ & 5.8$^{\circ}$ $\pm$ 0.07$^{\circ}$ & 11.49$^{\circ}$ $\pm$ 0.08$^{\circ}$ & 8.24$^{\circ}$ $\pm$ 0.07$^{\circ}$ & 66.08 \\ 
q = 0.3        & 12.03$^{\circ}$ $\pm$ 0.06$^{\circ}$ & 11.95$^{\circ}$ $\pm$ 0.07$^{\circ}$ & 11.55$^{\circ}$ $\pm$ 0.06$^{\circ}$ & 10.4$^{\circ}$ $\pm$ 0.06$^{\circ}$ & 11.94$^{\circ}$ $\pm$ 0.06$^{\circ}$ & 11.05$^{\circ}$ $\pm$ 0.07$^{\circ}$ & 27.36\\
q = 0.5        & 11.04$^{\circ}$ $\pm$ 0.06$^{\circ}$ & 10.76$^{\circ}$ $\pm$ 0.05$^{\circ}$ & 10.62$^{\circ}$ $\pm$ 0.06$^{\circ}$ & 10.74$^{\circ}$ $\pm$ 0.05$^{\circ}$ & 10.42$^{\circ}$ $\pm$ 0.05$^{\circ}$ & 10.5$^{\circ}$ $\pm$ 0.05$^{\circ}$ & 46.58\\
\enddata
\label{table_pitch_model} 
\end{deluxetable*}
\end{rotatetable*}


\bibliography{ref.bib}
\bibliographystyle{aasjournal}



\end{document}